\documentstyle[12pt,epsf]{article}
\renewcommand{\bar}[1]{\overline{#1}}

\begin{document}

\begin{flushright}
UNIV. NAPLES DSF 29/2001
\\ CPT-2001/P.4205
\end{flushright}
\bigskip\bigskip
\begin{center}
{\Large \bf A statistical approach \\
for polarized parton distributions}
\vspace{22pt}

{\bf
Claude  Bourrely} and {\bf Jacques Soffer}\\
Centre de Physique Th\'eorique, CNRS-Luminy Case 907,\\
13288 Marseille Cedex 9, France.
\vskip 0.5cm
{\bf Franco Buccella}\\
Dipartimento di Scienze Fisiche, Universit\`a di Napoli,\\
Via Cintia, I-80126, Napoli
and INFN, Sezione di Napoli, Italy\\
\end{center}
\vspace{32pt}

\centerline {\large \bf Abstract}

A global next-to-leading order QCD analysis of unpolarized and polarized
deep-inelastic scattering data is performed with parton distributions 
constructed in a statistical physical picture of the nucleon. 
The chiral properties of QCD lead to strong relations between quarks and 
antiquarks distributions and the importance of the Pauli exclusion principle
is also emphasized. We obtain a good description, in a broad range
of $x$ and $Q^2$, of all measured structure functions in terms of
very few free parameters. We stress the fact that at RHIC-BNL the ratio
of the unpolarized cross sections for the production of $W^+$ and $W^-$
in $pp$ collisions, will directly probe the behavior of the 
$\bar d(x) / \bar u(x)$ ratio for $x \geq 0.2$, a definite and important
test for the statistical model. Finally, we give specific predictions for
various helicity asymmetries for the $W^{\pm}, Z$ production in $pp$
collisions at high energies, which will be measured with forthcoming 
experiments at RHIC-BNL and are sensitive tests of the statistical model for 
$\Delta \bar u(x)$ and $\Delta \bar d(x)$.

\newpage
\section{Introduction}

Deep-inelastic scattering (DIS) of leptons on hadrons has been extensively
studied, over the last twenty years or so, both theoretically 
and experimentally.
The principal goals of this physics program were, first to elucidate the
internal proton structure, in terms of parton distributions, 
and more recently to test perturbative Quantum Chromodynamics
(QCD), which generalizes the parton model. For the unpolarized structure
functions, the advent of the HERA physics program gives us access 
to a broader kinematic range than fixed targets experiments, in $x$ down to a 
few $10^{-5}$ and in $Q^2$ up to several $10^4 \mbox{GeV}^2$, which allows 
testing perturbative QCD to next-to-leading order (NLO). 
As a result, the unpolarized light quarks ($u,d$) distributions are fairly 
well determined. Moreover, the data exhibit a clear evidence for a
flavor-asymmetric light sea, {\it i.e.} $\bar d > \bar u$, which can be
understood in terms of the Pauli exclusion principle, based on the fact 
that the proton contains two $u$ quarks and only one $d$ quark \cite{Pauli}. 
Larger uncertainties still persist for the gluon ($G$) and the heavy quarks 
($s,c$) distributions. From the more restricted amount of data on polarized 
structure functions, the corresponding polarized gluon and $s$ quark
distributions ($\Delta G, \Delta s$) are badly constrained and we just begin to
uncover a flavor asymmetry, for the corresponding polarized light sea, namely  
$\Delta \bar u \neq \Delta \bar d$. 
Whereas the signs of the polarized light quarks distributions are 
essentially well established, $\Delta u > 0$ and $\Delta d < 0$, 
this is not the case for $\Delta \bar u $ and $\Delta \bar d $.
The objective of this paper is to construct a complete set of polarized parton
(all flavor quarks, antiquarks and gluon) distributions and, in particular, 
we will try to clarify this last point on the polarized light sea.

The polarized parton distributions (PPD) of the nucleon have been extensively
studied in the last few years \cite{GRSV96,XXX-XXX} and in most models, 
the PPD are constructed from a set of unpolarized parton distributions, 
previously determined, from unpolarized DIS data. For example for each quark 
flavor $q_i(x)$, the corresponding $\Delta q_i(x)$ is taken 
(at the input energy scale) such that
\begin{equation}
\Delta q_i(x) = a_i(x)\cdot q_i(x)~,
\label{1}
\end{equation}
where $a_i(x)$ is a simple polynomial which has to be determined from the
polarized DIS data.
A similar procedure is used for antiquarks and gluons. As a result, the full
determination of all unpolarized and polarized parton distributions 
involves a large number of free parameters, say around 20-25, which obviously 
shows a lack of simplicity. In addition, most of these models
do not provide a flavor separation for the antiquarks $\bar q_i(x)$ and
consequently for $\Delta \bar q_i(x)$. 
However, there are recent attempts to make this flavor separation,
either using semi-inclusive polarized DIS data \cite{DDF00} or by means of a
flavor-symmetry breaking \cite{GRSV01}. 
Our motivation for this work is to use the statistical approach to
build up : $q_i$, $\Delta q_i$, $\bar q_i$, $\Delta \bar q_i$, $G$ and $\Delta
G$, by means of a very small number of free parameters. 
A flavor separation for the unpolarized and polarized light sea
is automatically achieved in a way dictated by our approach.

The paper is organized as follows. In Section 2, we review the main points of
our approach and we describe our method to determine the free parameters 
of the PPD with the set of experimental data we have used. 
In Section 3, we show the results obtained for the unpolarized DIS
structure functions $F_2^{p,d}(x,Q^2)$ and $xF_3^{\nu N}(x,Q^2)$ in a wide
kinematic range, compared with the world data. 
We show the prediction of the ratio of unpolarized $W^+$ and $W^-$ cross
section at RHIC-BNL, which is sensitive to the $\bar d(x) / \bar u(x)$ ratio, 
a challenging question for the statistical approach.
Section 4 is devoted to the polarized DIS structure functions
$g_1^{p,d,n}(x,Q^2)$. In Section 5, we give our predictions for single and
double helicity asymmetries for the heavy gauge boson production 
($W^{\pm}, Z$) in $pp$ collisions at high energies, which are sensitive 
to $\Delta \bar u $ and $\Delta \bar d$ and will be tested
with forthcoming experiments at RHIC-BNL. We give our concluding remarks in
Section 6.

\section{Basic procedure for the construction of the PPD in the statistical
approach}

In the statistical approach the nucleon is viewed as a gas of massless partons 
(quarks, antiquarks, gluons) in equilibrium at a given temperature in a 
finite size volume. 
Like in our earlier works on the subject \cite{B2M2ST,BSa,BSb}, we propose to 
use a simple description of the parton distributions $p(x)$, at an input 
energy  scale $Q_0^2$, proportional to
\begin{equation}
[\exp[(x - X_{0p})/{\bar x}] \pm 1]^{-1}~,
\label{2}
\end{equation}
the {\it plus} sign for quarks and antiquarks, corresponds to a Fermi-Dirac
distribution and the {\it minus} sign for gluons, corresponds 
to a Bose-Einstein distribution.
Here $X_{0p}$ is a constant which plays the role of the {\it thermodynamical
potential} of the parton {\it p} and $\bar x$ 
is the {\it universal temperature}, which is the same for all partons.
Since quarks carry a spin-1/2, it is natural to consider that the basic
distributions are $q_i^{\pm}(x)$, corresponding to a quark of flavor 
{\it i} and helicity parallel or antiparallel to the nucleon helicity.
This is the way we will proceed.
Clearly one has $q_i= q_i^{+} + q_i^{-}$ and $ \Delta q_i= q_i^{+} - q_i^{-}$
and similarly for antiquarks and gluons.

We want to recall that the statistical model of the nucleon has been
extensively studied in early and more recent papers in the literature 
\cite{Stat, Bha}, but in these works at variance with
our approach, the statistical picture is first considered in the nucleon rest
frame, which is then boosted to the infinite-momentum frame.

{}From the chiral structure of QCD, we have two important properties which
allow to relate quark and antiquark distributions and to restrict 
the gluon distribution \cite{BSb,Bha}:

- The potential of a quark $q_i^{h}$ of helicity {\it h} is opposite to the
potential of the corresponding antiquark $\bar q_i^{-h}$ of helicity {\it -h}
\begin{equation}
X_{0q}^h=-X_{0\bar q}^{-h}~.
\label{3}
\end{equation}

- The potential of the gluon $G$ is zero
\begin{equation}
X_{0G}=0~.
\label{4}
\end{equation}

{}From well established features of the $u$ and $d$ quark distributions
extracted from DIS data, we anticipate some simple relations 
between the potentials:

- $u(x)$ dominates over $d(x)$, therefore one can expect
$X^{+}_{0u} + X^{-}_{0u} > X^{+}_{0d} + X^{-}_{0d}$

- $\Delta u(x) > 0$, therefore $X_{0u}^+ > X_{0u}^-$

- $\Delta d(x) < 0$, therefore $X_{0d}^- > X_{0d}^+$~.

So we expect $X_{0u}^+$ to be the largest thermodynamical potential and
$X_{0d}^+$ the smallest one.
In fact, as we will see from the discussion below, we have the following
ordering
\begin{equation}
X_{0u}^+ > X_{0d}^- \sim  X_{0u}^- > X_{0d}^+ ~.
\label{5}
\end{equation}
Eq. (\ref{5}) is consistent with the previous determinations of the
potentials \cite{B2M2ST}, including the one with dimensional values
in the rest system \cite{Bha}.
By using Eq. ({3}), this ordering leads immediately to some important
consequences for antiquarks, namely

i) $\bar d(x) > \bar u(x)$, the flavor symmetry breaking which also follows
from the Pauli exclusion principle, as recalled above. 
This was already confirmed by the violation of the Gottfried sum rule 
\cite{Gott,NMC}.

ii) $\Delta \bar u(x) > 0$ and  $\Delta \bar d(x) < 0$, which remain to be
checked and this will be done in hadronic collisions at RHIC-BNL 
(see Section 6).

Note that since $\bar u^+(x) \sim \bar d^+(x)$, we have
\begin{equation}
\Delta \bar u(x) - \Delta \bar d(x) \sim  \bar d(x) -  \bar u(x) ~,
\label{6}
\end{equation}
so the flavor symmetry breaking is almost the same for unpolarized and
polarized distributions.

\begin{figure}
\begin{center}
\leavevmode {\epsfysize= 14.cm \epsffile{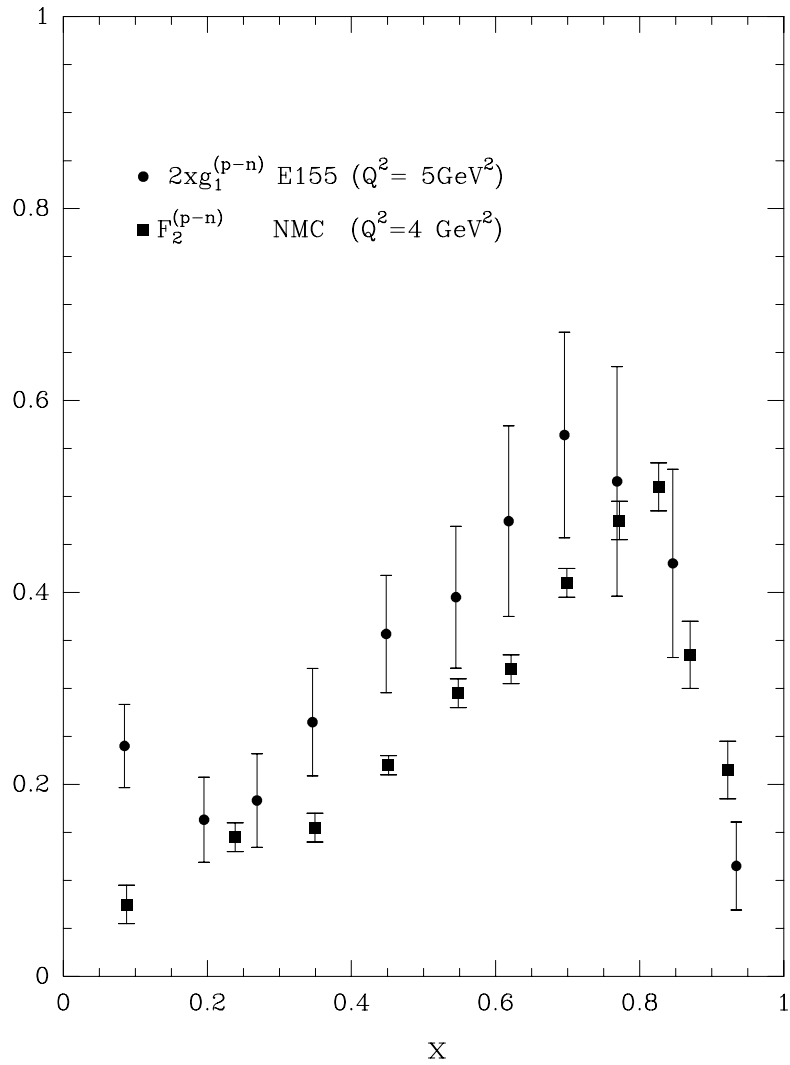}}
\end{center}
\caption[*]{\baselineskip 1pt
The isovector structure functions $ 2x g_1^{(p-n)}(x)$ and $F_2^{(p-n)}(x)$. 
Data is taken from Refs.~ \cite{NMC,E155}.
}\label{Fig1}
\end{figure}

Let us now come back to the ordering in Eq. (5) to justify it.
We consider the isovector contributions to the structure functions $g_1$ and
$F_2$, which are the differences on proton and neutron targets. 
In the QCD parton model they read
\begin{equation}
2x g_1^{(p-n)}(x,Q^2) = \frac{1}{3} x [ (\Delta u +\Delta \bar u)(x,Q^2)
-(\Delta d +\Delta \bar d)(x,Q^2) ]
\otimes \Delta C_{NS}(x,Q^2)~,
\label{7}
\end{equation}
and
\begin{equation}
F_2^{(p-n)}(x,Q^2) = \frac{1}{3} x [ ( u + \bar u)(x,Q^2)) - (d + \bar
d)(x,Q^2)) ] \otimes C_{NS}(x,Q^2)~,
\label{8}
\end{equation}
where $\Delta C_{NS}(x,Q^2)$ and $C_{NS}(x,Q^2)$ denote the spin-dependent and
spin-independent perturbative QCD coefficients \cite{GRSV96}. 
Since they differ only in a non-negligible way for very
small $x$, say $x\leq 0.05$, we see that $ 2x g_1^{(p-n)} - F_2^{(p-n)} $ is
only sensitive to the helicity {\it minus} components of the {\it u} 
and {\it d} quark distributions so we get
\begin{equation}
[2x g_1^{(p-n)} - F_2^{(p-n)}](x,Q^2) \sim -\frac{2}{3}
[ (u^{-}- d^{-})(x,Q^2)) + (\bar u^{-}- \bar d^{-})(x,Q^2))]
\otimes C_{NS}(x,Q^2)~.
\label{9}
\end{equation}

At this stage it is instructive to look at the available data shown in Fig. 1.
We notice that these two functions have very similar shapes and 
their difference
is small and mainly positive, except perhaps for large $x$. In order to try to
identify the origin of this experimental fact, let us look at the 
integrals of these functions divided by $x$. 
The first one is twice the Bjorken sum rule \cite{Bj}, for
which the best world estimate is $I_{Bj}=0.176 \pm 0.005 \pm 0.007$ \cite{E155}
and the second one is the Gottfried sum rule \cite{Gott} whose value is
$I_G=0.235 \pm 0.026$ \cite{NMC}.
As a result using Eq. (\ref{9}) one obtains, say for $Q^2=5\mbox{GeV}^2$,
\begin{equation}
\int_{0}^{1} dx [(d^{-}(x) - u^{-}(x)) + (\bar d^{-}(x) - \bar u^{-}(x))] =
0.175 \pm 0.06~.
\label{10}
\end{equation}
Now from the NMC result on the Gottfried sum rule one has
\begin{equation}
\int_{0}^{1} dx [ \bar d(x) - \bar u(x) ] = 0.16 \pm 0.03~.
\label{11}
\end{equation}
By comparing these two results we can assume, to a good approximation, the
following relation for the helicity {\it minus} distributions
\begin{equation}
d^{-}(x) = u^{-}(x) ~.
\label{12}
\end{equation}
It follows from our procedure to construct antiquark distributions from quark
distributions described above (see Eq. (3)), that we have automatically for 
the helicity {\it plus} antiquark distributions
\begin{equation}
\bar d^{+}(x) = \bar u^{+}(x)~,
\label{13}
\end{equation}
which makes Eqs. ({10}) and ({11}) perfectly compatible. Indeed, as we will 
see below, Eq. ({12}) is rather well satisfied in the final determination of 
the distributions, after fitting the data.

Let us now complete the description of our parametrization. As stated above,
the essential ingredient for quarks and antiquarks is a Fermi-Dirac 
distribution, as shown in Eq. ({2}), but we expect this piece to die 
out in the small {\it x} region, so we have to multiply it
by a factor $AX_{0q}^h x^b$, where $ b > 0$. In addition to {\it A}, a
flavor and helicity independent normalization constant, we have 
introduced the factor $X_{0q}^h$ which is needed to get a good
description of the data. It is not required by the simple Fermi-Dirac
expression but, due to the ordering in Eq. ({5}), it will 
secure the correlation between the shape of a given distribution and its 
first moment \cite{B2M2ST,FBJS}. It is also in agreement with what has been
found from data for the second and third moments of the valence partons
\cite{BDPRS}. The small {\it x} region
is characterized by a rapid rise as $ x \rightarrow 0$ of the distribution,
which should be dominated by a universal diffractive term, 
flavor and helicity independent, coming from the Pomeron universality. 
Therefore we must add a term of the form
${\tilde{A}x^{\tilde b}} / [\exp(x /\bar x)+1]$, where $\tilde b < 0$
and $\tilde A$ is a normalization constant.
So for the light quarks {\it q = u, d} of helicity $h =\pm$, at the input
energy scale $Q_0^2=4\mbox{GeV}^2$, we take
\begin{equation}
xq^{h}(x,Q_0^2)= \frac{A X_{0q}^h x^b}{\exp[(x-X_{0q}^h)/{\bar x}]+1} +
\frac{{\tilde A} x^{\tilde b}}{\exp(x/{\bar x})+1}~,
\label{14}
\end{equation}
and similarly for the light antiquarks
\begin{equation}
x\bar q^{h}(x,Q_0^2)= \frac{\bar A (X_{0q}^{-h})^{-1}
x^{2b}}{\exp[(x+X_{0q}^{-h})/{\bar x}]+1}
+\frac{{\tilde A}x^{\tilde b}}{\exp(x/{\bar x})+1}~.
\label{15}
\end{equation}
Here we take $2b$ for the power of $x$ and not $b$ as for quarks, 
an assumption we will try to justify later.
For the strange quarks and antiquarks, {\it s} and  $\bar s $, given our poor
knowledge on both unpolarized and polarized distributions, 
we take the particular choice
\begin{equation}
xs(x,Q_0^2)=x \bar s(x,Q_0^2)= \frac{1}{4}[x \bar u (x,Q_0^2) + x \bar
d(x,Q_0^2)]~,
\label{16}
\end{equation}
and
\begin{equation}
x\Delta s(x,Q_0^2)=x \Delta \bar s(x,Q_0^2)= \frac{1}{3}[x \Delta \bar d
(x,Q_0^2) -x \Delta \bar u(x,Q_0^2)]~.
\label{17}
\end{equation}
This particular choice gives rise to a large negative 
$\Delta s(x,Q_0^2)$ and we will
come back to it below, in the discussion of our results (see Section 4).
The charm quarks {\it c }, both unpolarized and polarized, are set to zero at
$Q_0^2=4\mbox{GeV}^2$.
Finally concerning the gluon distribution, as indicated above, we use a
Bose-Einstein expression given by
\begin{equation}
xG(x,Q_0^2)=\frac{A_Gx^{b_G}}{\exp(x/{\bar x})-1}~,
\label{18}
\end{equation}
with a vanishing potential and the same temperature $\bar x$.
This choice is consistent with the idea that hadrons, in the DIS regime, 
are black body cavities for the color fields. 
It is also reasonable to assume that for very small {\it x}, $xG(x,Q_0^2)$ 
has the same behavior as $x\bar q(x,Q_0^2)$, so we will take $b_G=1+\tilde b$. 
Since the normalization constant $A_G$ is determined from the momentum
sum rule, our gluon distribution has no free parameter. For the sake of
completeness, we also need to specify the polarized gluon distribution and 
we take the particular choice
\begin{equation}
x \Delta G(x,Q_0^2) = 0~,
\label{19}
\end{equation}
consistently with Eq. ({4}).
As usual, the valence contributions are defined as $q_{val}= q - \bar q$, so
{\it A} and $\bar A$ are determined using the normalization of 
$u_{val}(x)$ and $d_{val}(x)$, whose first moments are respectively 2 and 1.

To summarize our parametrization involves a total of {\it eight} free 
parameters
\begin{equation}
\bar x~, X_{0u}^+~, X_{0u}^-~, X_{0d}^-~, X_{0d}^+ ~,b~, 
\tilde{b}~ \mbox{and} {}~ \tilde{A}~.
\label{20}
\end{equation}
In order to determine these parameters, we use a fitting procedure on a
selection of 233 data points at $Q^2$ values, as close  as possible to our 
input energy scale $Q_0^2=4\mbox{GeV}^2$  and the $\chi^2$ 
value we obtain is 322.
For unpolarized DIS, we have considered $F_2^p(x,Q^2)$ from NMC, BCDMS, E665
and ZEUS, $F_2^d(x,Q^2)$ from NMC, E665 and $xF_3^{\nu N}(x,Q^2)$ from CCFR
\cite{NMC},\cite{ALL91}-\cite{QUI}. 
For polarized DIS we have considered $g_1^{p,d,n}(x,Q^2)$ 
from SMC, E154 and  E155 \cite{E155,ADE,ABE54}.
The {\it five} free parameters, {\it one} temperature and {\it four} 
potentials, which determine the Fermi-Dirac functions at the input
energy scale $Q_0^2=4\mbox{GeV}^2$ are
\begin{equation}
\bar x=0.09907,~ X_{0u}^+=0.46128,~ X_{0u}^-=0.29766,~
X_{0d}^-=0.30174~\mbox{and}~ X_{0d}^+=0.22775~.
\label{21}
\end{equation}
{}From the above discussion which led us to Eqs. (12) and (13), we observe that
the fit yields $X_{0u}^-\sim X_{0d}^-$, in agreement with our expectations.
We show in Figs. 2,~3  the Fermi-Dirac functions Eq. (2) for light quarks and
antiquarks, respectively. 
They exhibit a flat behavior in the small $x$ region, at variance 
with the rising trend of the parton distributions, which ought to be described 
by the universal diffractive term (see Eqs. (14,~15)). It is also interesting
to make one more observation from the above values of these potentials. 
It turns out
that if we impose the following simple relations $X_{0d}^-=X_{0u}^-$, 
$X_{0u}^+=3/2X_{0u}^-$ and
$X_{0d}^+=3/4X_{0u}^-$, we can get an equally good fit of the data with 
$X_{0u}^-=0.30549$.
This choice reduces the number of free parameters for the potentials 
from four to one, but we cannot justify it and it might be fortuitous.

For the remaining {\it three} free parameters $b$~, 
$\tilde{b}$ and $\tilde{A}$, the fit gives the values
\begin{equation}
b=0.40962,~ \tilde{b}=-0.25347~ \mbox{and} ~ \tilde{A}=0.08318~.
\label{22}
\end{equation}
Finally the parameters $A$, $\bar A$ and $A_G$ determined by normalization
conditions and momentum sum rule, have the following values
\begin{equation}
A=1.74938,~ {\bar A}=1.90801~ \mbox{and} ~ A_G=14.27535~.
\label{23}
\end{equation}
We show in Figs. 4,~5, the different helicity components of the 
light quarks and
antiquarks, respectively, at $Q^2=20\mbox{GeV}^2$, after a NLO evolution. 
They all have the same rising behavior in the small $x$-region, which is 
driven by the universal diffractive term. We note that after NLO evolution, we
still have $u^- \sim d^-$ and $\bar u^+ \sim \bar d^+$.
In Fig. 6, we display the $x$-shapes of the full set of
unpolarized parton distributions, where one sees a non zero $c$ quark
distribution generated by the $Q^2$ evolution.
Note that $u^{-} \sim d^{-}$ implies
\begin{equation}
u(x) - d(x) \sim \Delta u(x) - \Delta d(x)
\label{23b}
\end{equation}
at least up to $Q^2 \sim 20\mbox{GeV}^2$. This is a specific feature of our
quark distributions which is not fulfilled in most parametrizations
\cite{XXX-XXX}.
In order to compute the evolved distributions, we have used the following
method: all parton distributions and splitting functions are 
decomposed in terms of Chebyshev polynomials, then the DGLAP equations at NLO
are solved by a Runge-Kutta method \cite{kwie} or by a semianalytic method
\cite{santorel} (the renormalization scheme adopted is $\bar {MS}$
with $\Lambda_{\bar {MS}}[n_f=3] = 300\mbox{MeV})$.
We have checked that both methods are consistent, within the
numerical accuracy. All the results shown in the figures are calculated at NLO.

Let us now comment on the values obtained for some of these
parameters. First, it is interesting to note that $A$ and $\bar A$, the
normalizations of quarks and antiquarks, come close to each other. 
Next, concerning the power of $x$ for which we took, $b$ for quarks and 
$2b$ for antiquarks, let us try to understand this
fact and the value of $b$. The universal diffractive term, involving $\tilde b$
and $\tilde A$, is absent in the physical quantity $x\bar d(x) - x\bar u(x)$,
which was extracted recently in Ref. \cite{E866} (see Section 3 for 
further comments on these data).
This quantity rises in the region $0.01 \leq x \leq 0.2$, whereas the antiquark
Fermi-Dirac functions decrease (see Fig. 3). 
This rising behavior is simply obtained if one multiplies
these functions by $x^{0.8}$. Another physical quantity which does not contain
the universal diffractive term is $xF_3^{\nu N}(x)$, because 
it involves the difference $q_i(x) - \bar q_i(x)$.
In the region $0.01 \leq x \leq 0.2$, quarks dominate over antiquarks 
(see Fig. 6) and we know that $xF_3^{\nu N}(x)$ rises. 
This behavior is obtained if one multiplies the quark
Fermi-Dirac function by $x^{0.4}$, in agreement with Eq. (22).

In the next section we will compare our calculations obtained using these
parton distributions, with the existing experimental world data, 
for the unpolarized DIS structure functions $F_2^{p,d}(x,Q^2)$ 
and $xF_3^{\nu N}(x,Q^2)$, in a wide kinematic range.
We will also show the prediction of the ratio of unpolarized $W^+$ and $W^-$
cross sections at RHIC-BNL, which is sensitive to  the $\bar d(x) / \bar u(x)$ 
ratio, a challenging question for the statistical approach.
\begin{figure}
\begin{center}
\leavevmode {\epsfysize= 13.cm \epsffile{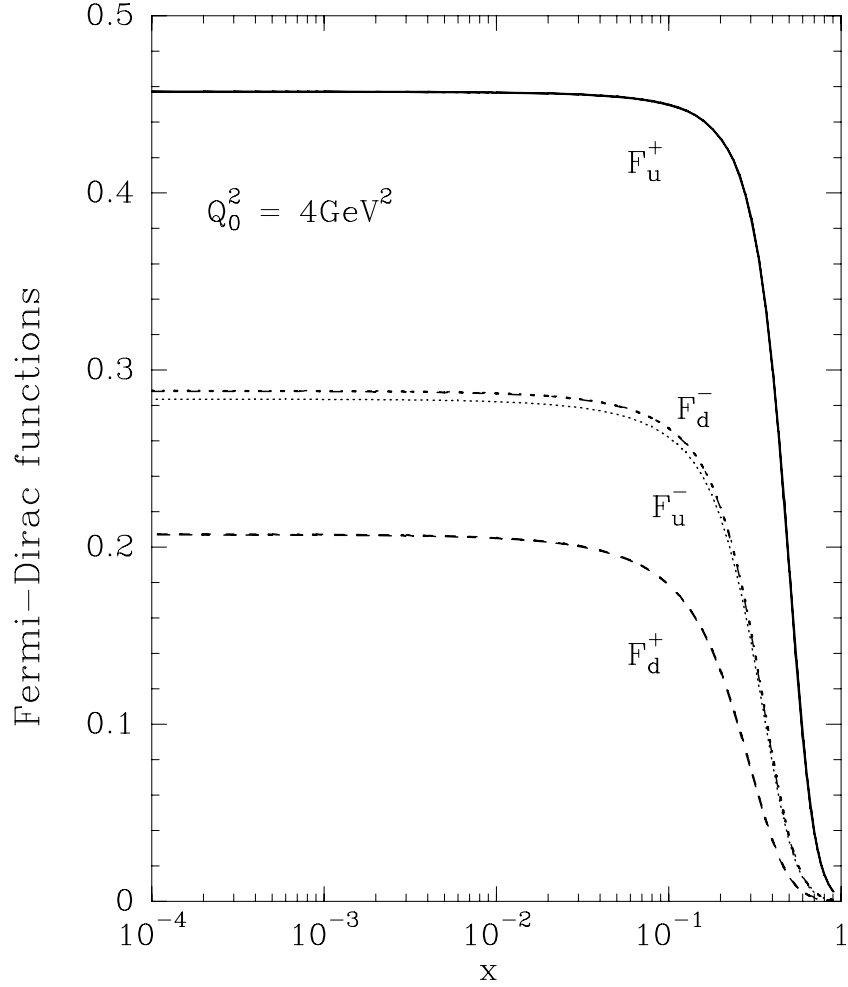}}
\end{center}
\caption[*]{\baselineskip 1pt
The Fermi-Dirac functions for quarks $F_q^h = X^h_{0q}/
(\exp[(x-X_{0q}^h)/{\bar x}]+1)$
at the input energy scale $Q_0^2=4\mbox{GeV}^2$, as a function of $x$.
}\label{Fig2}
\end{figure}

\clearpage
\newpage
\begin{figure}
\begin{center}
\leavevmode{\epsfysize= 14.cm \epsffile{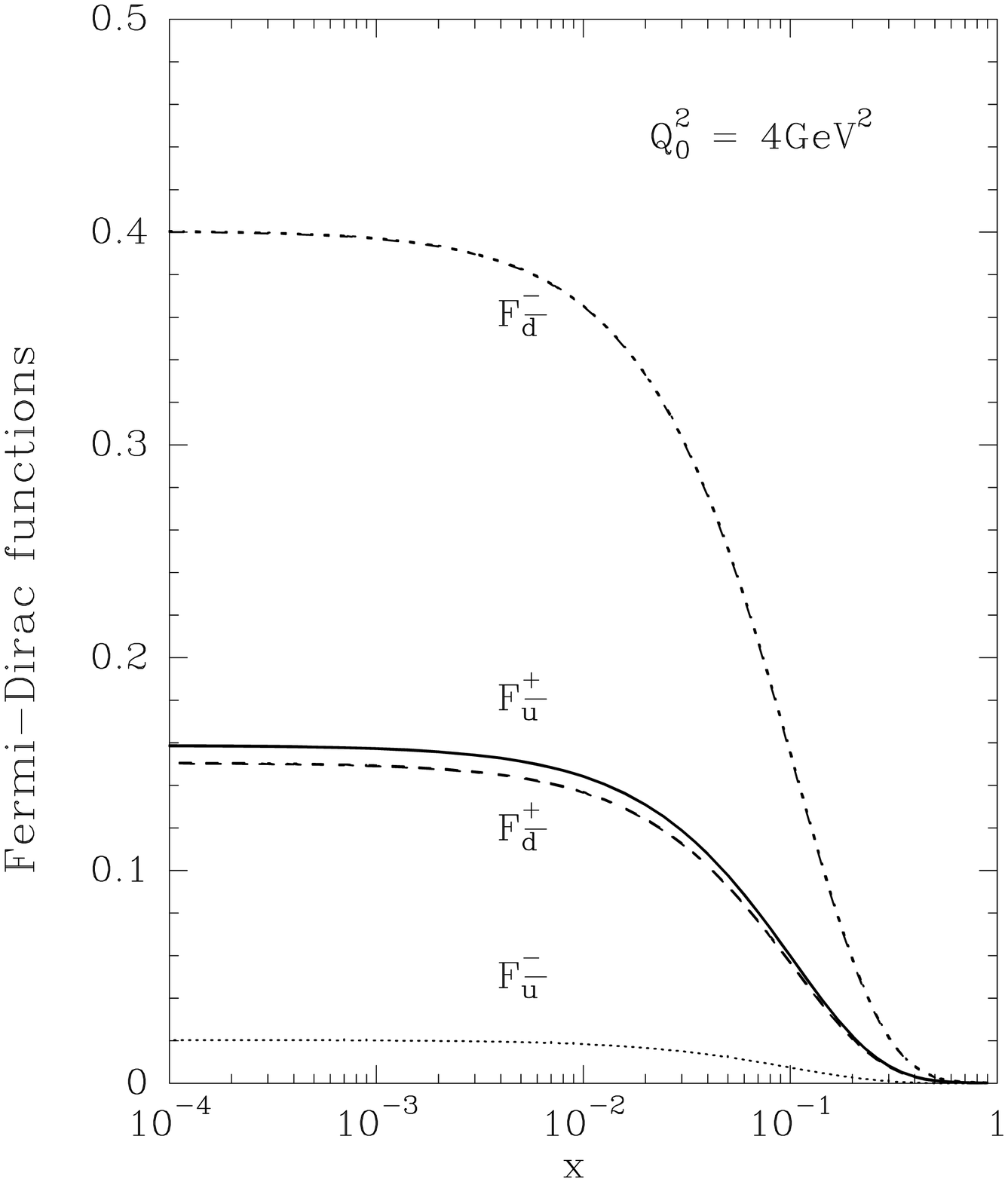}}
\end{center}
\caption[*]{\baselineskip 1pt
The Fermi-Dirac functions for antiquarks $F_{\bar q}^h = 1/X^h_{0\bar
q}(\exp[(x+X_{0\bar q}^h)/{\bar x}]+1)$
at the input energy scale $Q_0^2=4\mbox{GeV}^2$, as a function of $x$.
}\label{Fig3}
\end{figure}

\clearpage
\newpage
\begin{figure}
\begin{center}
\leavevmode {\epsfysize= 14cm \epsffile{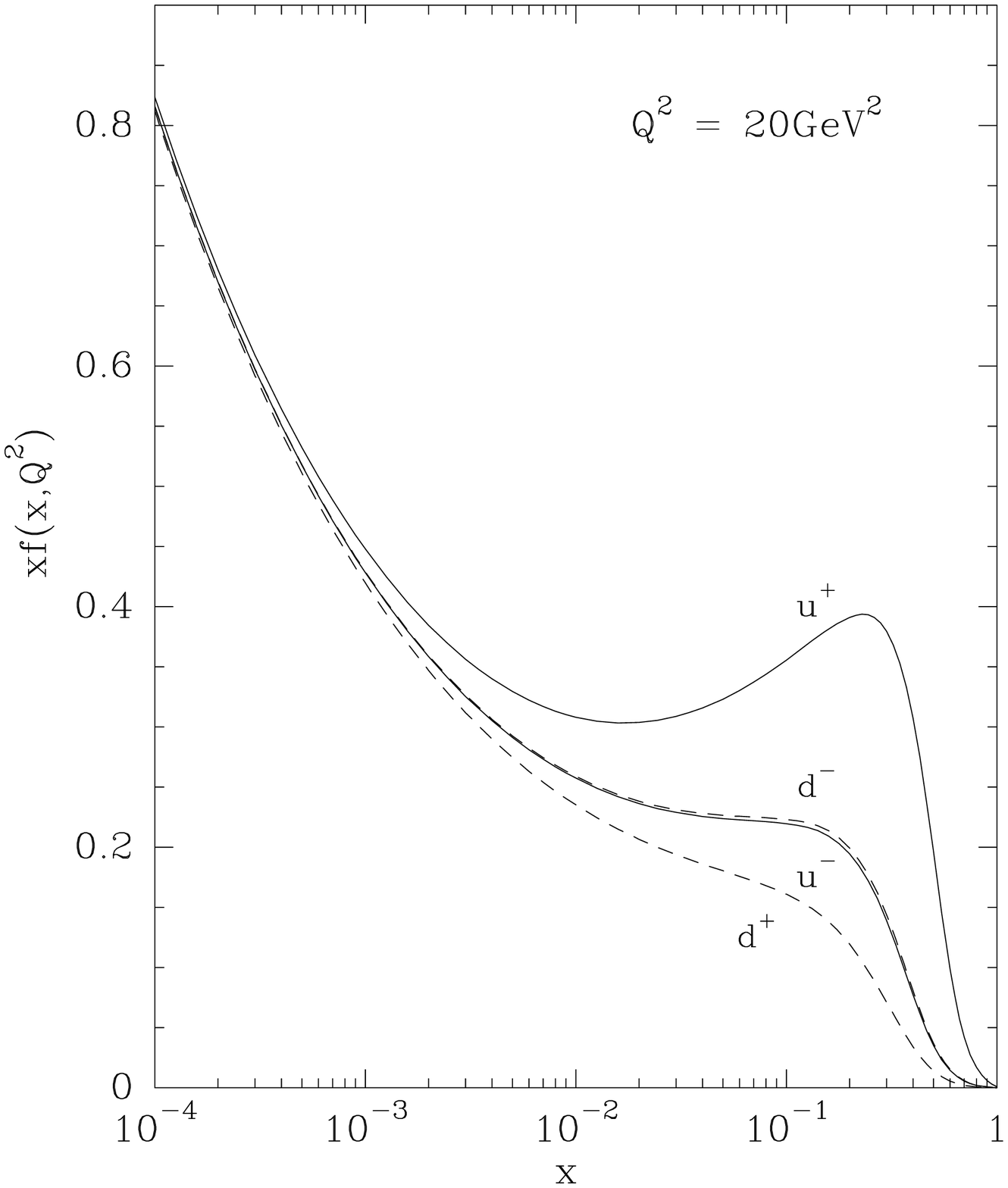}}
\end{center}
\caption[*]{\baselineskip 1pt
The different helicity components of the light quark distributions after NLO
evolution, at $Q^2=20\mbox{GeV}^2$, as a function of $x$.
}\label{Fig4}
\end{figure}

\clearpage
\newpage
\begin{figure}
\begin{center}
\leavevmode {\epsfysize= 14cm \epsffile{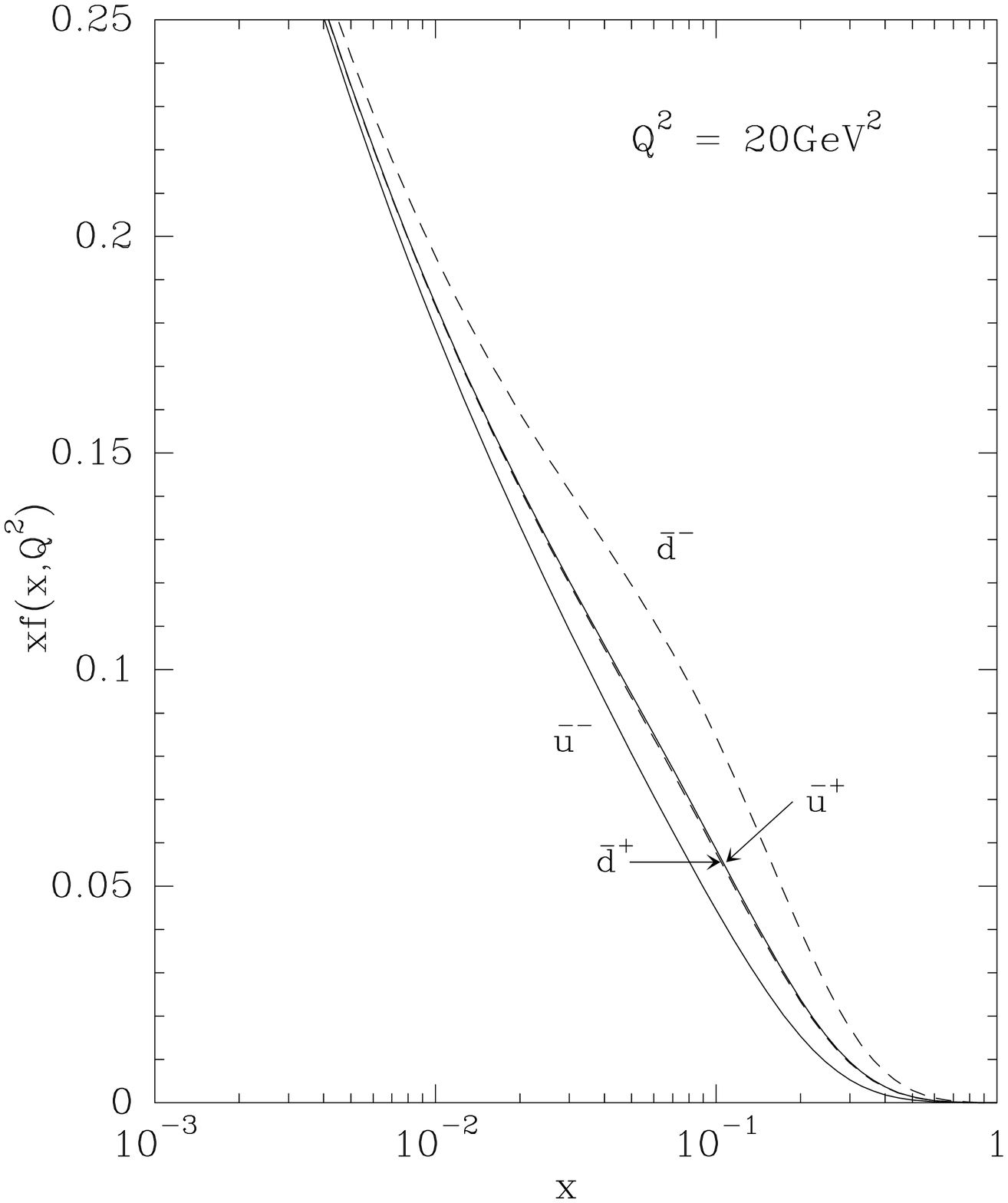}}
\end{center}
\caption[*]{\baselineskip 1pt
The different helicity components of the light antiquark distributions after
NLO evolution, at $Q^2=20\mbox{GeV}^2$, as a function of $x$.}
\label{Fig5}
\end{figure}

\clearpage
\newpage
\begin{figure}
\begin{center}
\leavevmode {\epsfysize= 14cm \epsffile{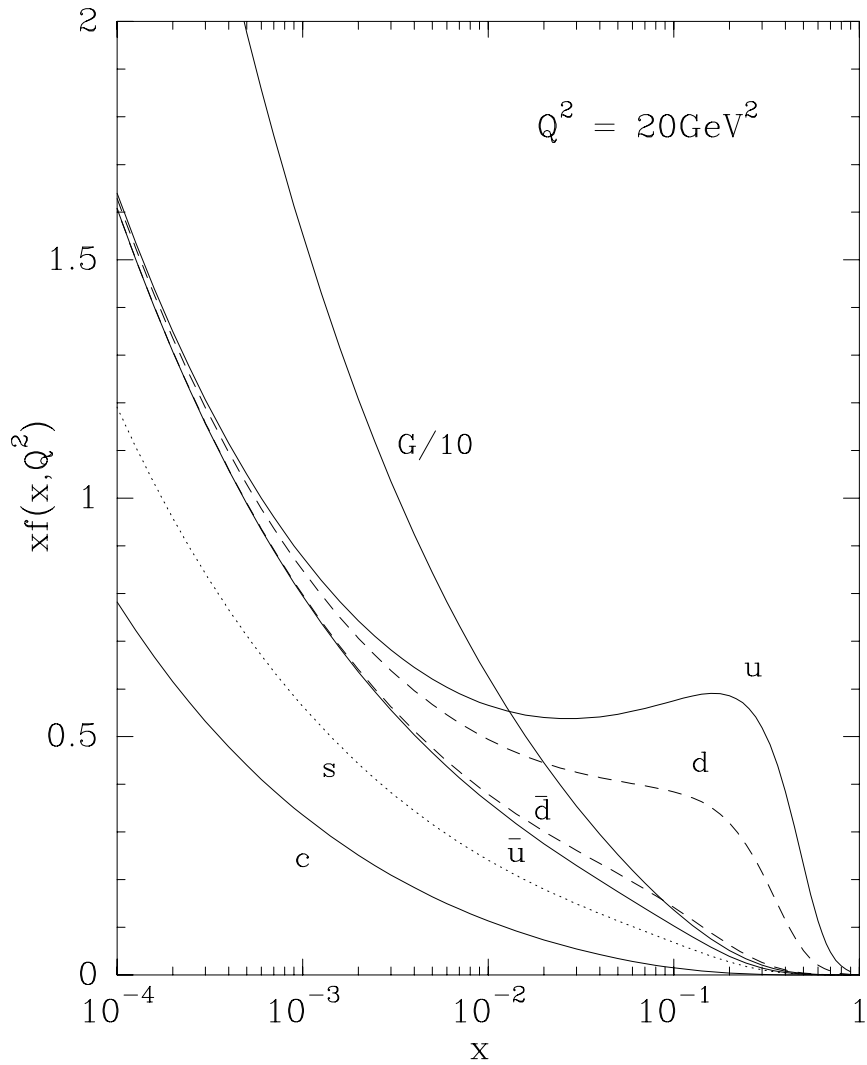}}
\end{center}
\caption[*]{\baselineskip 1pt
The different unpolarized parton distributions ($f=$$u$,~$d$,
~$\bar u$,~$\bar d$,~$s$,~$c$ and $G$) after NLO evolution, at 
$Q^2=20\mbox{GeV}^2$, as a function of $x$.}
\label{Fig6}
\end{figure}

\clearpage
\newpage

\section{Experimental tests for unpolarized DIS}

We first consider $\mu p$ and $ep$ DIS for which several experiments have
yielded a large number of data points on the structure function
$F_2^p(x,Q^2)$.
We have compared our calculations with fixed target
measurements NMC, BCDMS and E665, which cover a rather
limited kinematic region in $Q^2$ and also with the data at HERA from the H1
and ZEUS Collaborations.
These last data cover a very large $Q^2$ range, up to $Q^2=10^4\mbox{GeV}^2$
or so and probe the very low $x$ region which is dominated
by the rising behavior of the universal diffractive term.

The comparison of our results with the data is shown in Figs. 7,~8,~9 and 10.
We notice in Fig.~ 7 that the very low $Q^2$ range accessible by E665,
requires a downwards $Q^2$ evolution which is achieved successfully
down to $Q^2=1\mbox{GeV}^2$ or so, but we fail to reproduce the data much
below than that. The H1 and ZEUS data are fairly well described
as shown in Figs. 8 and 9, respectively. Finally, we present a compilation of
the  data, including NMC and BCDMS, in Fig. 10 which is in very good 
agreement with our theoretical curves.
{}From measurements over a large $Q^2$ range, it is possible to improve the
determination of the gluon density by analyzing the scaling 
violations. On Fig. 11 we see that $xG(x,Q^2)$ exhibits a fast rising behavior
in the low $x$ region, which is fairly consistent with our simple 
parametrization (see Eq. (18)). Note that, $xG(x,Q^2)$ is predicted 
to increase with $Q^2$ as shown in the figure. Next we consider $F_2^d(x,Q^2)$
obtained on a deuterium fixed target from NMC, BCDMS and E665 data. 
The comparison of these very accurate data with our results is shown on 
Figs. 12,~13 and 14, respectively. 
The agreement is also excellent, except for the very low $Q^2$ 
region of E665, as for the proton case. 
Finally the high statistics $\nu N$ DIS data from CCFR allows to extract the 
$xF_3^{\nu N}(x,Q^2)$ structure function, which is successfully 
compared to our results on Fig. 15. 

To complete our tests of the unpolarized parton distributions, we must come 
back to the important question of the flavor asymmetry of the light
antiquarks. Our determination of $\bar u(x,Q^2)$ and
$\bar d(x,Q^2)$ is perfectly consistent with the violation of the Gottfried
sum rule, for which we found $I_G= 0.2493$ for $Q^2=4\mbox{GeV}^2$.
Nevertheless there remains an open problem with the $x$ distribution
of the ratio $\bar d/\bar u$ for $x \geq 0.2$.
According to the Pauli principle this ratio should be above 1 for any value of
$x$. However, recently the E866/NuSea Collaboration \cite{E866} has
released the final results corresponding to the analysis of their full
data set of Drell-Yan yields from an 800 GeV/c proton beam on hydrogen
and deuterium targets and they obtain the ratio, for $Q^2=54\mbox{GeV}^2$, 
$\bar d/\bar u$ shown in Fig.~16. 
Although the errors are rather large in the high $x$ region,
the statistical approach disagrees with the trend of the data.
Clearly by increasing the number of free parameters, it
is possible to build up a scenario which leads to the drop off of
this ratio for $x\geq 0.2$.
For example this was achieved in Ref. \cite{Sassot}, as shown 
by the dashed curve in Fig. 16. There is no such freedom in the statistical
approach, since quark and antiquark distributions are strongly related.
One way to clarify the situation is, either to improve the statistical
accuracy on the Drell-Yan yields which seems rather unlikely, or to call
for the measurement of another observable sensitive to
$\bar u(x)$ and  $\bar d(x)$.
One possibility is the ratio of the unpolarized cross sections for the
production of $W^+$ and $W^-$ in $pp$ collisions, which will directly probe
the behavior of the $\bar d(x) / \bar u(x)$ ratio.
Let us recall that if we denote 
$R_W(y)=(d\sigma^{W^+}/dy)/(d\sigma^{W^-}/dy)$, where $y$ is the
$W$ rapidity,  we have \cite{BSc} at the lowest order
\begin{equation}
R_W(y,M_W^2)= \frac{u(x_a,M_W^2) \bar d(x_b,M_W^2) + \bar d(x_a,M_W^2)
u(x_b,M_W^2)}{d(x_a,M_W^2) \bar u(x_b,M_W^2)
+ \bar u(x_a,M_W^2) d(x_b,M_W^2)}~,
\label{24}
\end{equation}
where $x_a=\sqrt{\tau}e^y$, $x_b=\sqrt{\tau}e^{-y}$ and $\tau=M_W^2/s$.
This ratio $R_W$, such that $R_W(y)=R_W(-y)$, is accessible 
with a good precision at RHIC-BNL \cite{BSSW} and at 
$\sqrt s = 500\mbox{GeV}$ for $y=0$, we have $x_a=x_b=0.16$.
So $R_W(0,M_W^2)$ probes the $\bar d(x) / \bar u(x)$ ratio at $x=0.16$. 
Much above this $x$ value, the accuracy of Ref. \cite{E866} becomes poor.
In Fig. 17 we compare the results for $R_W$ using two different
calculations.
In both cases we take the $u$ and $d$ quark distributions
obtained from the present analysis, but first we use the $\bar u$ and
$\bar d$ distributions of the statistical approach
(solid curve in Fig. 16) and second the $\bar u$ and $\bar d$ from
Ref. \cite{Sassot} (dashed curve in Fig. 16).
Notice that the energy scale $M_W^2$ is much higher than in the E866/NuSea
data, so one has to take into account the $Q^2$ evolution. 
At $\sqrt s = 200\mbox{GeV}$ for $y=0$, we have $x_a=x_b=0.40$ and, 
although the 
$W^{\pm}$  yield is smaller at this energy, the effect on $R_W(0,M_W^2)$ is
strongly enhanced, as seen in Fig. 17. This excellent test,
must be done in the near future.
 
\begin{figure}
\begin{center}
\leavevmode {\epsfysize= 14cm \epsffile{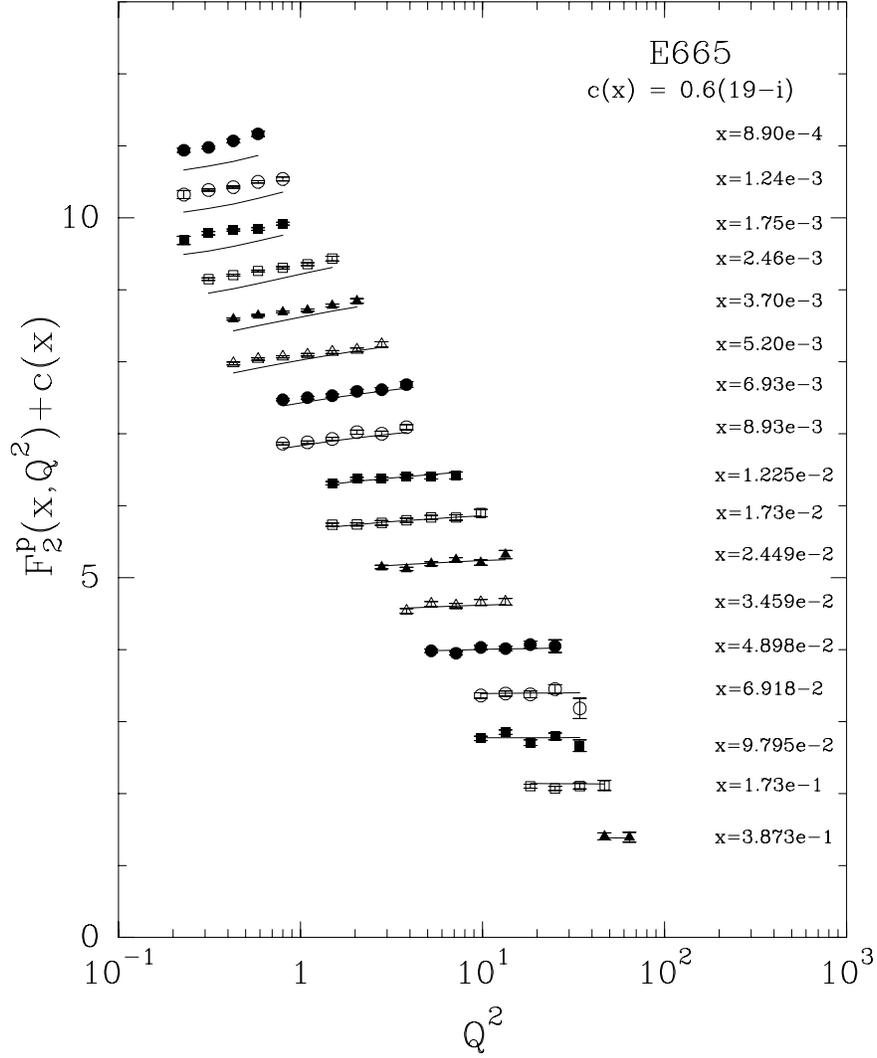}}
\end{center}
\caption[*]{\baselineskip 1pt
$F_2^p(x,Q^2)$ as function of $Q^2$ for fixed $x$, E665 data \cite{E665a}.
The function $c(x_i) = 0.6(19 -i)$, $i = 1$ corresponds to $x = 8.9~10^{-4}$.}
\label{Fig7}
\end{figure}
 
\begin{figure}
\begin{center}
\leavevmode {\epsfysize= 14cm \epsffile{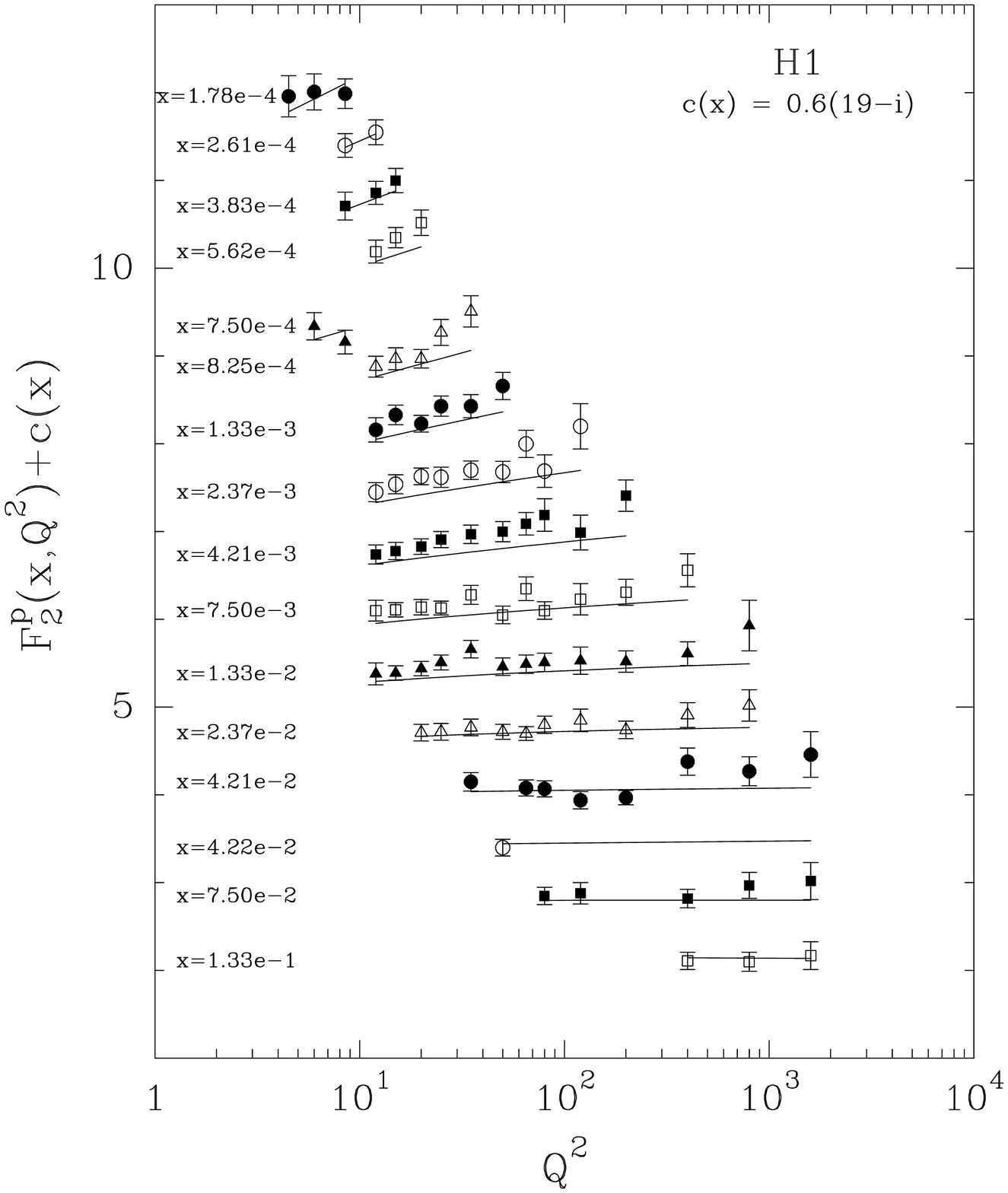}}
\end{center}
\caption[*]{\baselineskip 1pt
$F_2^p(x,Q^2)$ as function of $Q^2$ for fixed $x$, H1 data \cite{H1a, H1b}.
The function $c(x_i) = 0.6(19 -i)$, $i = 1$ corresponds to $x = 1.78~10^{-4}$.}
\label{Fig8}
\end{figure}
 
\begin{figure}
\begin{center}
\leavevmode {\epsfysize= 14cm \epsffile{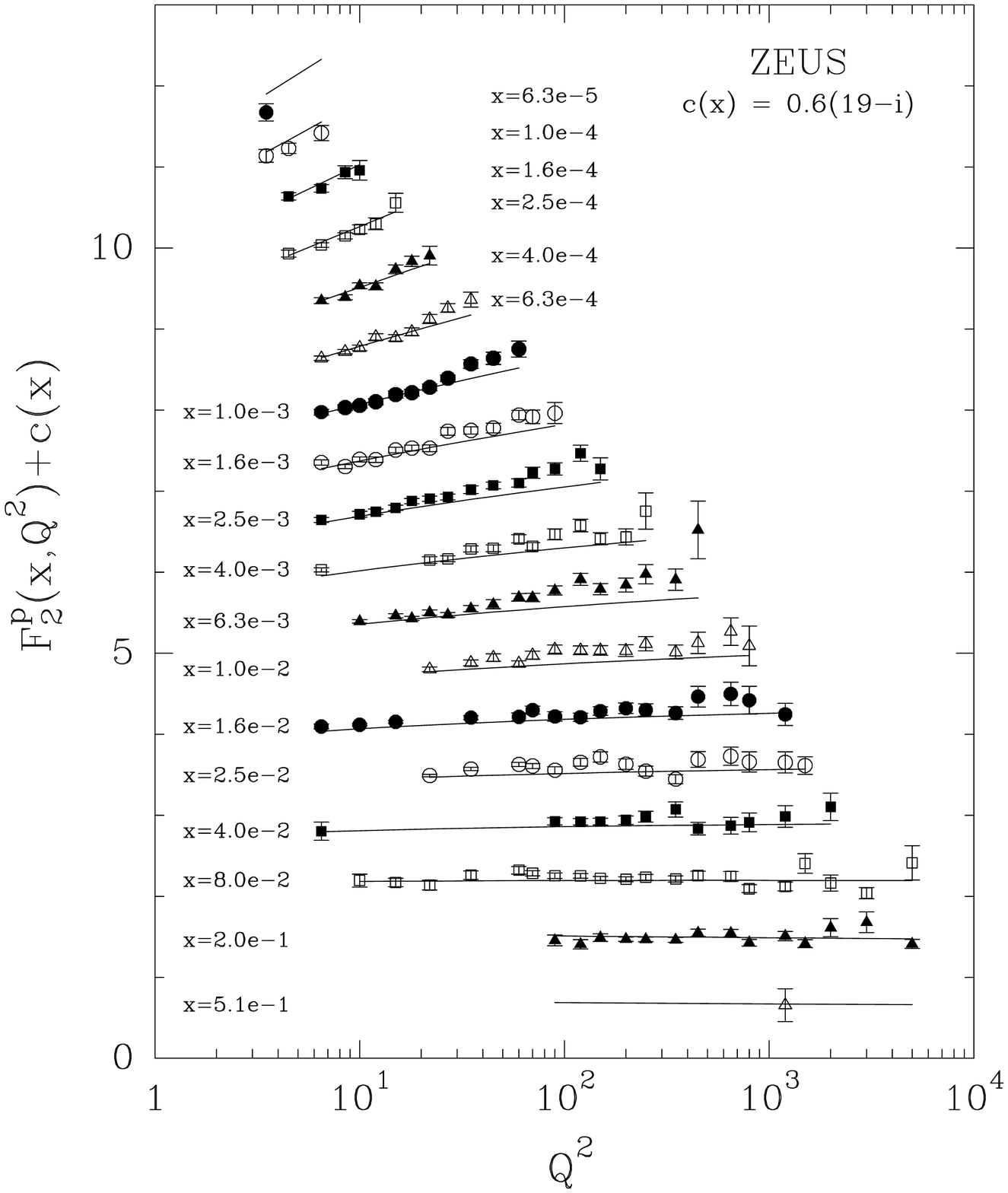}}
\end{center}
\caption[*]{\baselineskip 1pt
$F_2^p(x,Q^2)$ as function of $Q^2$ for fixed $x$, ZEUS data
\cite{ZEUSa,ZEUSb}.
The function $c(x_i) = 0.6(19 -i)$, $i = 1$ corresponds to $x = 6.3~10^{-5}$.}
\label{Fig9}
\end{figure}
 
\begin{figure}
\begin{center}
\leavevmode {\epsfysize= 14cm \epsffile{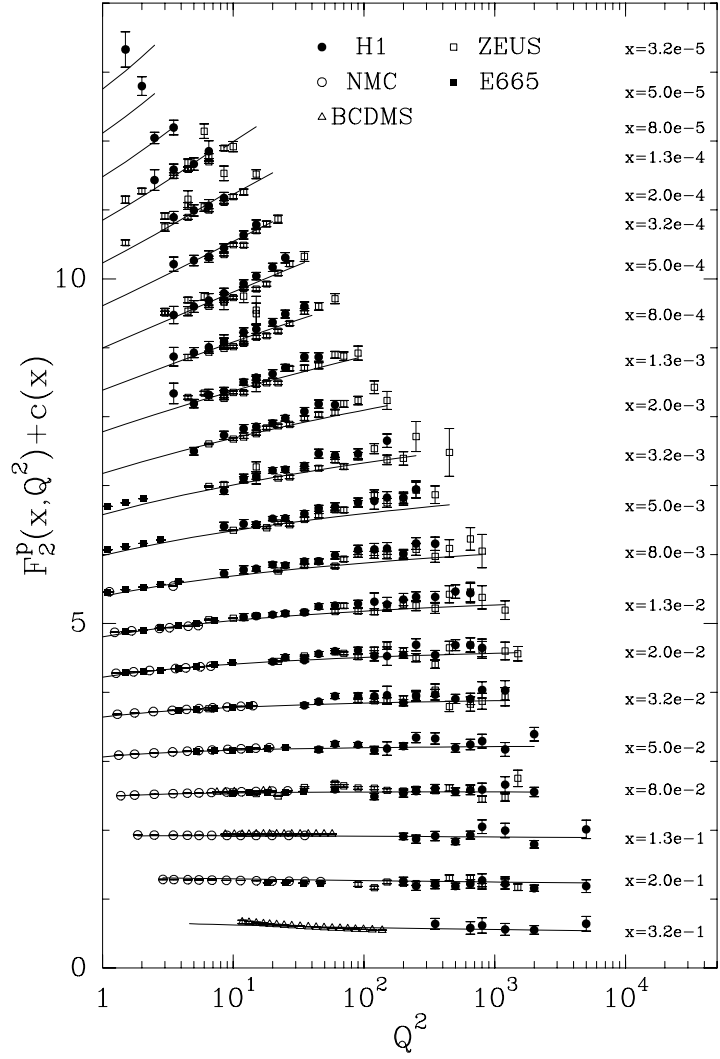}}
\end{center}
\caption[*]{\baselineskip 1pt
$F_2^p(x,Q^2)$ as function of $Q^2$ for fixed $x$,
$c(x) =0.6(i_x-0.4)$, $i_x = 1 \rightarrow x = 0.32$,
rebinned data H1, ZEUS, E665, NMC, BCDMS. (Presentation of data, courtesy of 
R. Voss).
}\label{Fig10}
\end{figure}

\begin{figure}
\begin{center}
\leavevmode{\epsfysize= 14.cm \epsffile{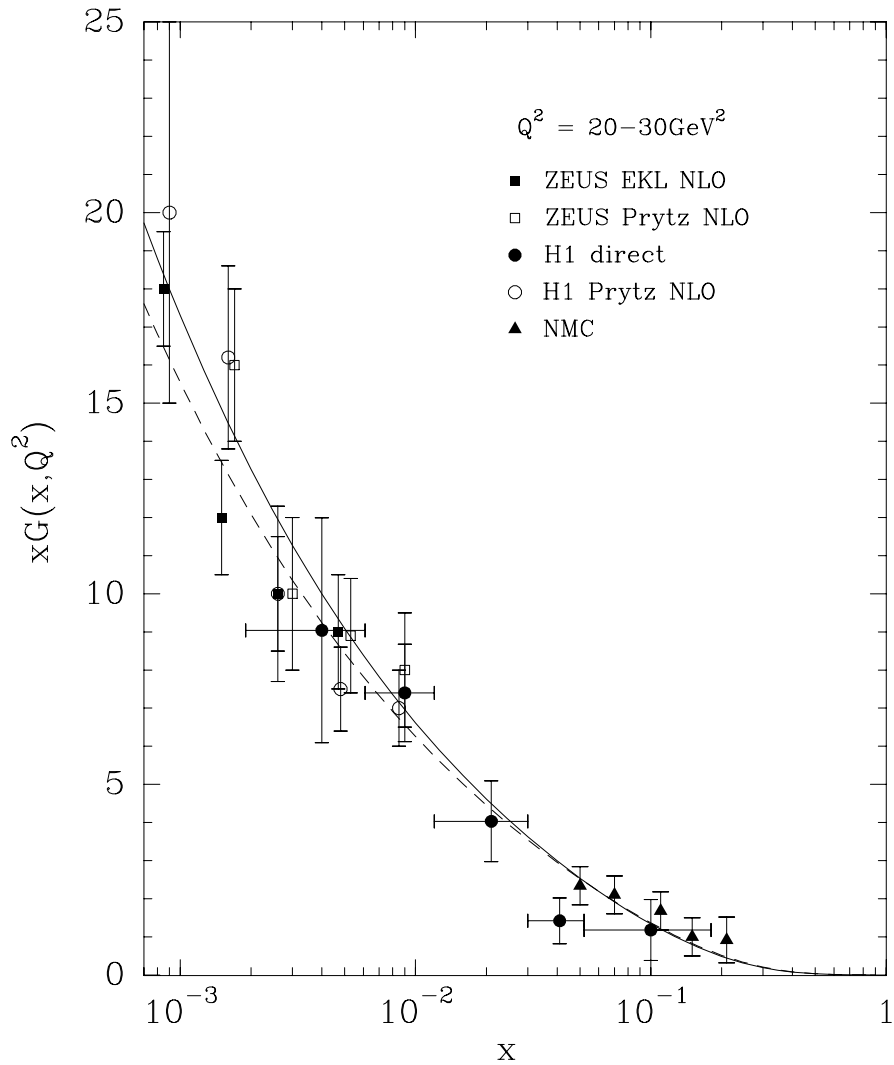}}
\end{center}
\caption[*]{\baselineskip 1pt
Comparison of $xG(x,Q^2)$ at $Q^2=20-30 \mbox{GeV}^2$ (dashed-solid)
with experimental determination from NMC \cite{ALL91}, 
H1 \cite{AID95} and ZEUS \cite{DER95}  experiments.}
\label{Fig11}
\end{figure}

\begin{figure}
\begin{center}
\leavevmode{\epsfysize= 14.cm \epsffile{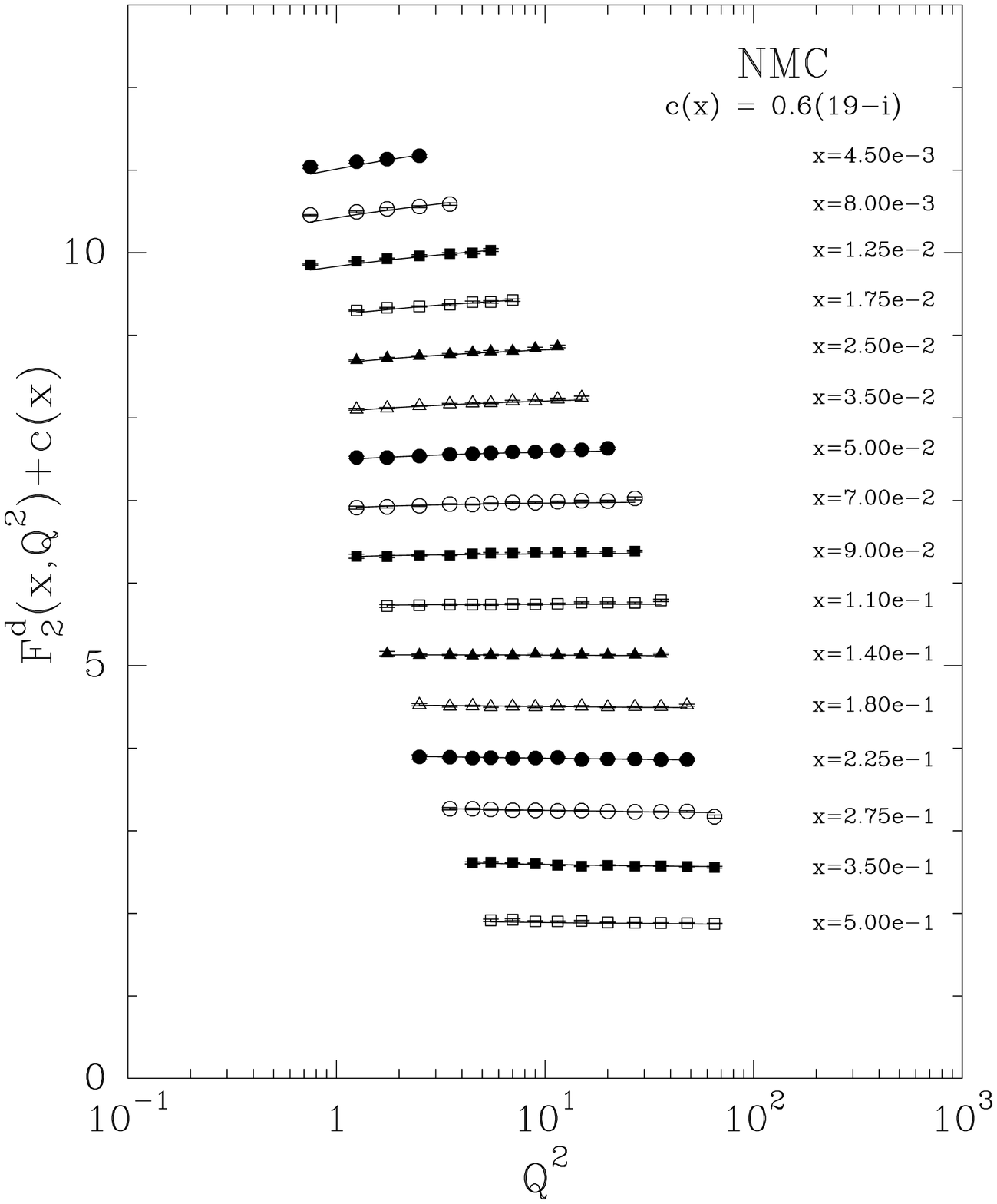}}
\end{center}
\caption[*]{\baselineskip 1pt
$F_2^d(x,Q^2)$ as function of $Q^2$ for fixed $x$, NMC data 
\cite{NMCa}. The function $c(x_i) = 0.6(19 -i)$, $i = 1$ 
corresponds to $x = 4.5~10^{-3}$.}
\label{Fig12}
\end{figure}

\begin{figure}
\begin{center}
\leavevmode {\epsfysize= 14cm \epsffile{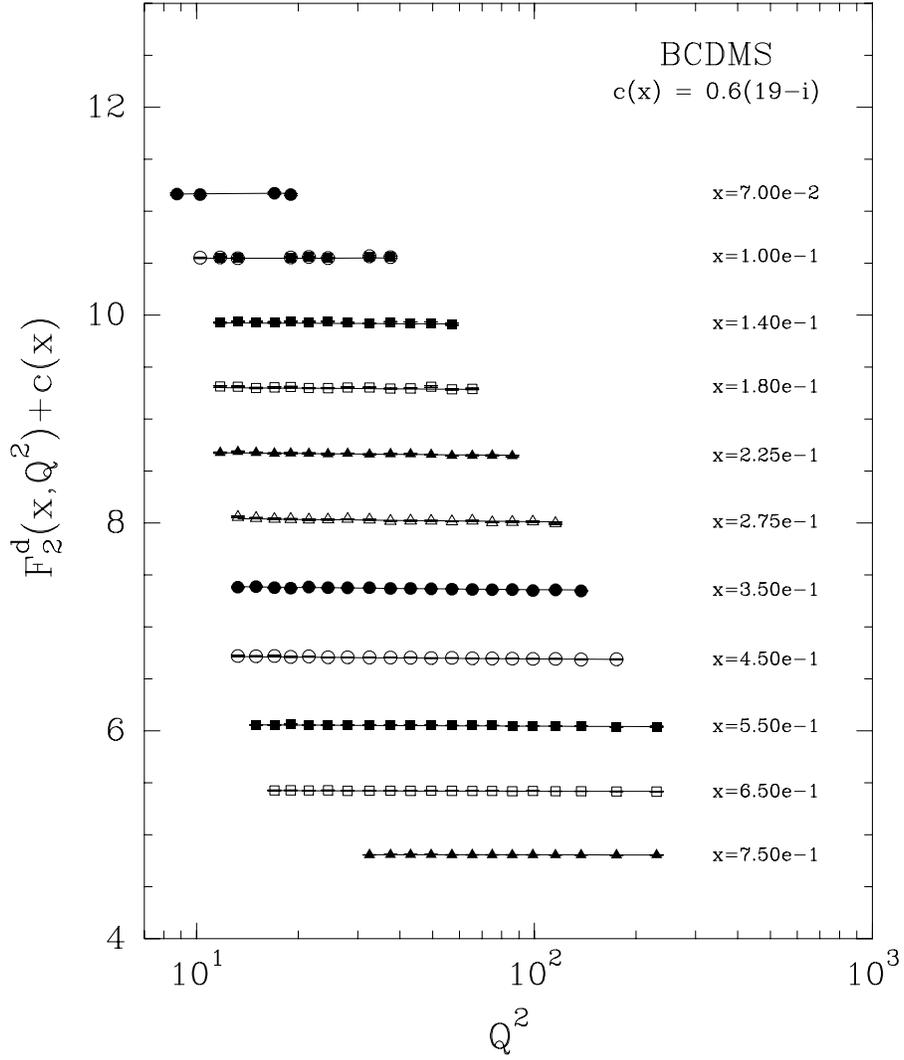}}
\end{center}
\caption[*]{\baselineskip 1pt
$F_2^d(x,Q^2)$ as function of $Q^2$ for fixed $x$, BCDMS data \cite{BCDMSc}.
The function $c(x_i) = 0.6(19 -i)$, $i = 1$ corresponds to $x = 7~10^{-2}$.}
\label{Fig13}
\end{figure}

\begin{figure}
\begin{center}
\leavevmode {\epsfysize= 14cm \epsffile{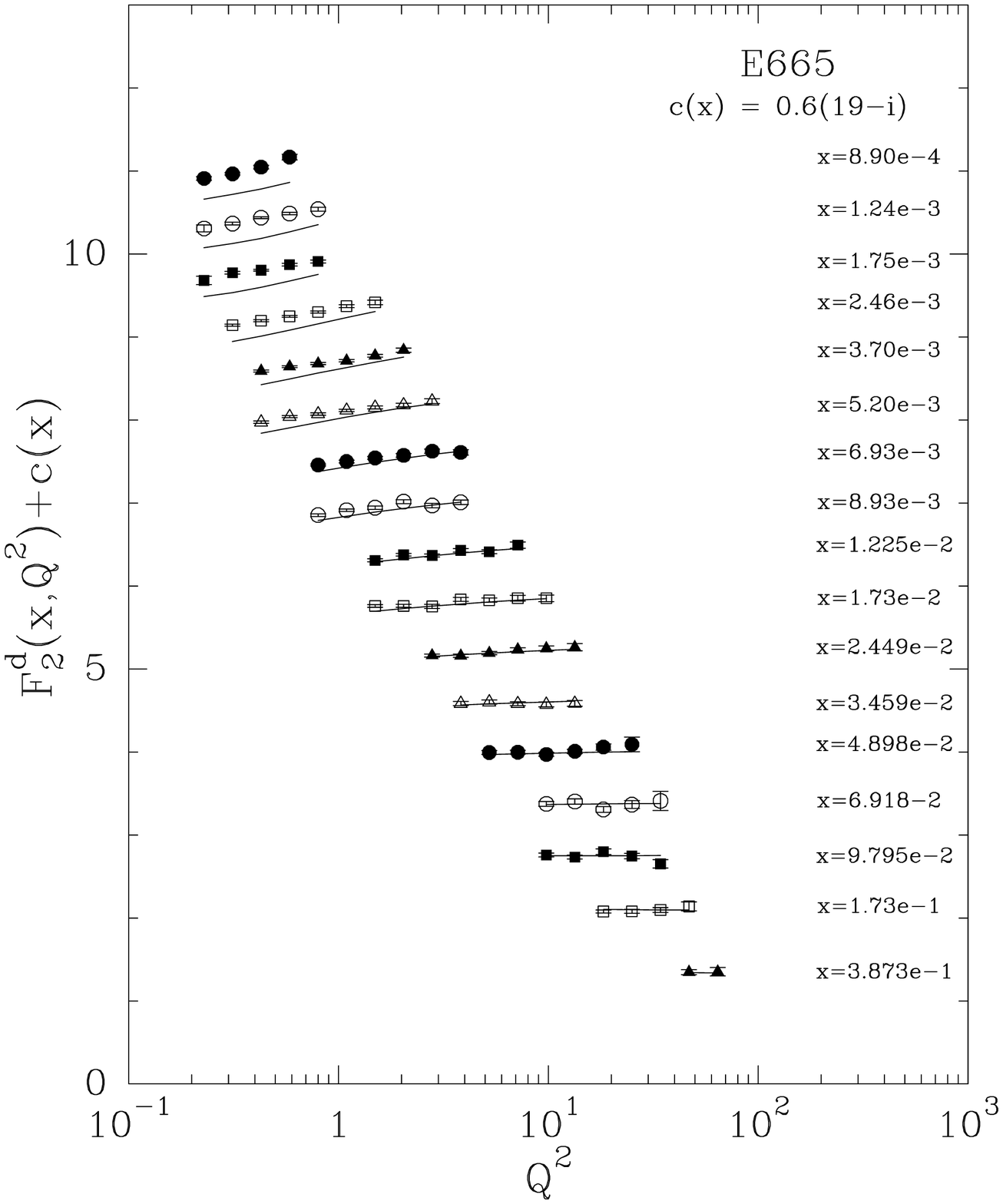}}
\end{center}
\caption[*]{\baselineskip 1pt
$F_2^d(x,Q^2)$ as function of $Q^2$ for fixed $x$, E665 data
\cite{E665a}.
The function $c(x_i) = 0.6(19 -i)$, $i = 1$ corresponds to $x = 8.9~10^{-4}$.}
\label{Fig14}
\end{figure}

\begin{figure}
\begin{center}
\leavevmode {\epsfysize= 14cm \epsffile{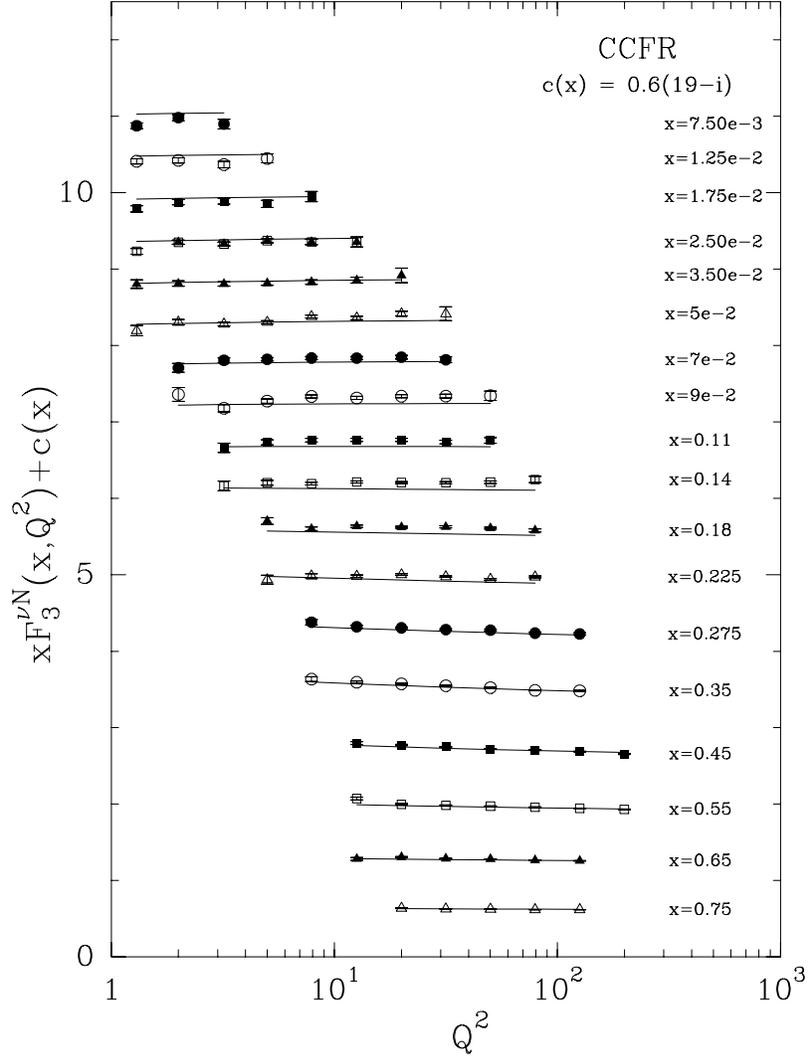}}
\end{center}
\caption[*]{\baselineskip 1pt
$xF_3^{\nu N}(x,Q^2)$ as function of $Q^2$ for fixed $x$, CCFR data
\cite{QUI}.
The function $c(x_i) = 0.6(19 -i)$, $i = 1$ corresponds to $x = 7.5~10^{-3}$.}
\label{Fig15}
\end{figure}

\begin{figure}
\begin{center}
\leavevmode {\epsfysize= 14cm \epsffile{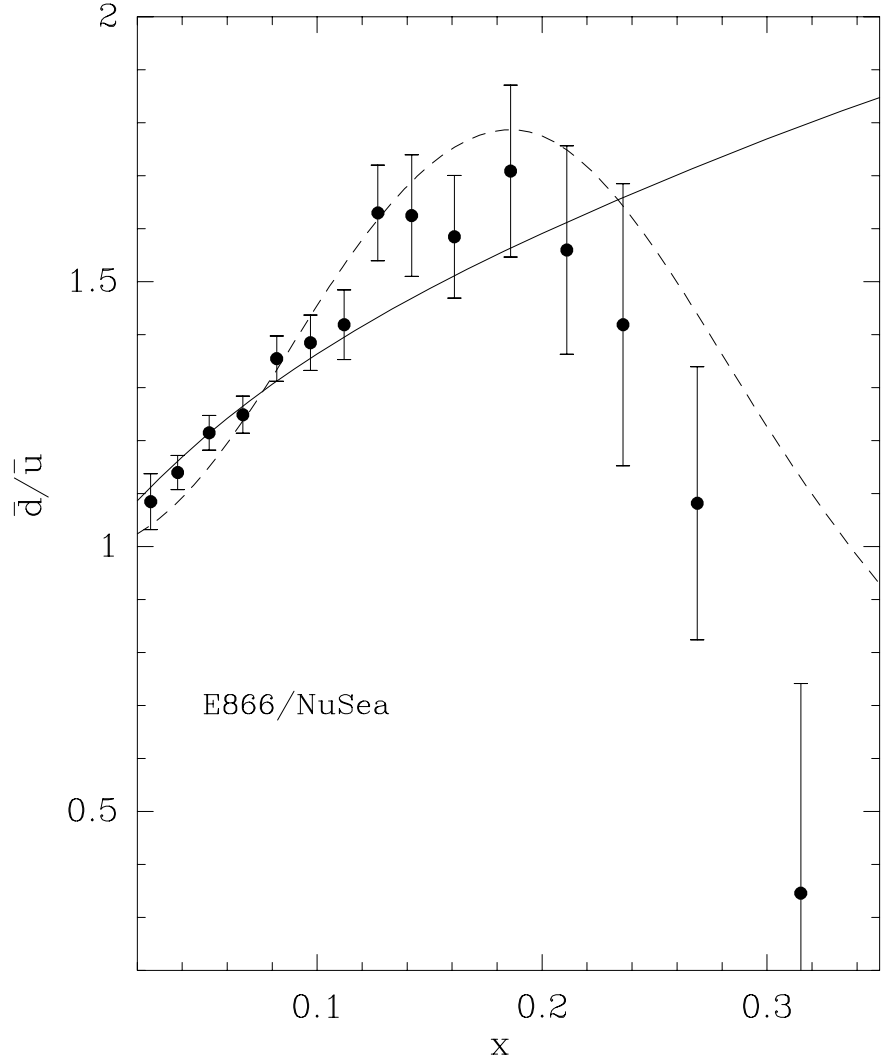}}
\end{center}
\caption[*]{\baselineskip 1pt
Comparison of the data on $\bar d / \bar u (x,Q^2)$  from E866/NuSea
at $Q^2=54\mbox{GeV}^2$
\cite{E866}, with the prediction of the statistical model (solid curve) 
and the set 1 of the parametrization proposed in Ref. \cite{Sassot}
(dashed curve).}
\label{Fig16}
\end{figure}

\begin{figure}
\begin{center}
\leavevmode {\epsfysize= 14cm \epsffile{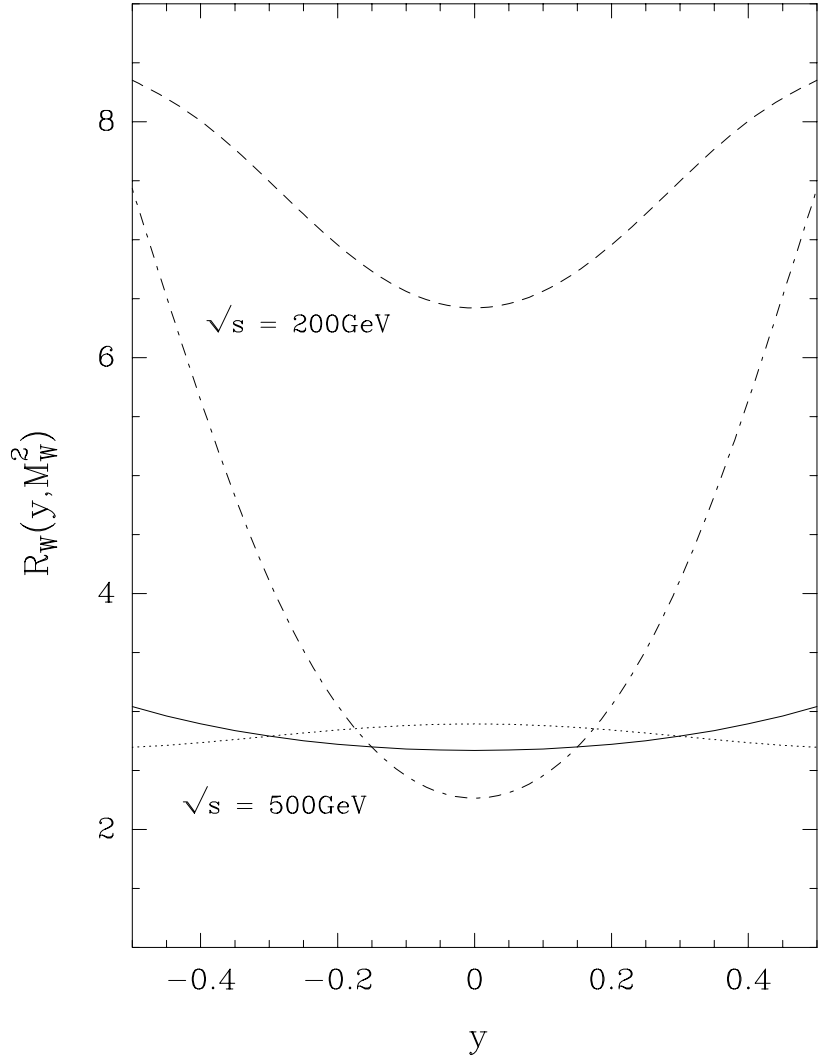}}
\end{center}
\caption[*]{\baselineskip 1pt
Theoretical calculations for the ratio $R_W(y,M_W^2)$ versus the $W$ rapidity,
at two RHIC-BNL energies. 
Solid curve ($\sqrt s = 500\mbox{GeV}$) and dashed curve 
($\sqrt s = 200\mbox{GeV}$) are the statistical model predictions. 
Dotted curve ($\sqrt s = 500\mbox{GeV}$) and dashed-dotted curve 
($\sqrt s = 200\mbox{GeV}$) 
are the predictions obtained using the $\bar d(x) / \bar u(x)$ ratio
from Ref. \cite{Sassot}.}
\label{Fig17}
\end{figure}

\clearpage
\newpage

\section{Experimental tests for polarized DIS}

Since our approach is based on the direct construction of the quark and
antiquark distributions of a given helicity  $q_i^{\pm}$ and $\bar q_i^{\pm}$, 
from the previous results we immediately obtained $\Delta q_i$ 
and $\Delta \bar q_i$ for each flavor.
We display in Fig. 18 these distributions $x\Delta f(x,Q^2)$ versus $x$, at
$Q^2 = 20\mbox{GeV}^2$, after a NLO evolution. 
As we mentioned earlier, we took $\Delta G =\Delta s = \Delta \bar s = 0$ 
at the input energy scale, but as shown in the Figure, this 
is no longer true after the $Q^2$ evolution.
We notice that the distributions are positive for $u$,~$\bar u$ and $G$ and
negative for $d$,~$\bar d$,~$s (\bar s)$ and $c$, which 
remain extremely small.
We have also checked that our $\Delta q_i(x)$ satisfy the positivity conditions
at the leading twist level obtained in Ref. \cite{ST}. Recently, the HERMES
Collaboration has presented new semi-inclusive data \cite{hermes} with greater
precision, which allows a good flavor separation of the light quarks 
and these data are shown in Fig. 19.
However from this data analysis one cannot achieve a flavor separation for the
antiquarks, so $\Delta q_s / q_s$ represents the polarization of 
sea quarks assuming flavor symmetry, {\it i.e.}
$\Delta q_s / q_s = \Delta \bar u / \bar u = \Delta \bar d / \bar d = \Delta s
/ s $.
Our model calculations give a very good description of the $u$ and $d$ quark
polarizations as shown in the Figure. 
The comparison between our three curves
$\Delta \bar u / \bar u$,~ $\Delta \bar d / \bar d $ and $ \Delta s / s$ and
the poor accuracy sea quark data does not allow one to draw any conclusion.

In Fig. 20 we show a data compilation of the polarized structure functions
$g_1^{p,d,n}(x,Q^2)$ from different current experiments on proton, 
deuterium and helium targets, evolved at a fixed value $Q^2 = 5\mbox{GeV}^2$. 
The $x$ dependence is in fair agreement with our results and we predict, in
the small $x$ region, a fast rising behavior for $g_1^p$ and a fast
decreasing behavior for $g_1^n$, due to the antiquark contributions. 
This cannot be tested so far, due to the lack of precise data. 
The $Q^2$ dependence for fixed $x$ values is displayed in Figs. 21 and 22.

Here we would like to comment on the choice we made for $\Delta s(x,Q^2)$
($\Delta \bar s$) in Eq. ({17}).
Clearly this polarized quark distribution is very badly known and we have
constrained its first moment by assuming the validity of the second Bjorken 
sum rule, namely
\begin{equation}
\Delta q_8 = \Delta u + \Delta \bar u + \Delta d + \Delta \bar d - 2(\Delta s +
\Delta \bar s) = 3F - D~,
\label{25}
\end{equation}
where $F$ and $D$ are the hyperon beta decay constants, so that $3F-D=0.579$.
At $Q^2 = 4\mbox{GeV}^2$, we have found for the following first moments, 
$\Delta u = 0.6678$, $\Delta \bar u = 0.0464$, $\Delta d = -0.2576$,
$\Delta \bar d= -0.0865$ and $\Delta s = \Delta \bar s = -0.0443$, so 
$\Delta q_8 = 0.547$. 
Notice that the first Bjorken sum rule is also very well satisfied, 
since we get $I_{Bj}=0.1764$. It is interesting to realize that
the contribution of the antiquarks to $I_{Bj}$, whose value in our case 
is 0.0221, is positive and
relatively large because $\Delta \bar u$ and $\Delta \bar d$ have the same 
signs as $\Delta u$ and $\Delta d$, respectively.

Finally we turn to the important issue of the large $x$ behavior of the
polarized quark distributions. 
This kinematic region has been poorly explored experimentally so far, but
there are different theoretical scenarios \cite{xto1}, when $x$ is near 1,
for the asymmetries $A_1^{p,d,n}(x,Q^2)$, measured in polarized DIS.
We recall the definition of the asymmetry $A_1(x,Q^2)$, namely
\begin{equation}
A_1(x,Q^2)= \frac{g_1(x,Q^2)}{F_2(x,Q^2)}
\frac{2x[1+R(x,Q^2)]}{[1+\gamma^2(x,Q^2)]}~,
\label{26}
\end{equation}
where $\gamma^2(x,Q^2)=4M^2x^2/Q^2$ and $R(x,Q^2)$ is the ratio between the
longitudinal and transverse photoabsorption cross sections. 
In the case where the $u$ quark dominates, we get
\begin{equation}
A_1 \sim \frac{\Delta u(x,Q^2)}{ u(x,Q^2)}
\frac{[1+R(x,Q^2)]}{[1+\gamma^2(x,Q^2)]}~.
\label{27}
\end{equation}
When $x \rightarrow 1$ for $Q^2 = 4\mbox{GeV}^2$, $R$ is of the order of 0.30
or less and $\gamma^2(x,Q^2)$ is close to 1, so 
$A_1 \sim 0.6 \Delta u(x)/u(x)$. It seems unlikely
to find $A_1 \rightarrow 1$, unless one lets $Q^2$ go to infinity or one 
violates positivity.
We show in Fig. 23 a compilation of the world data for $A_1^{p,n}(x,Q^2)$ at
$Q^2 = 4\mbox{GeV}^2$, with the results of our calculations up to $x=1$, 
 where we have $\Delta u/u =0.77$ , $\Delta d/d =-0.46$ and indeed we find
$A_1^{p,n}<1$.
This specific prediction should be confronted with the very accurate data 
on $A_1^n$ in the large $x$ region which is expected soon from
Jefferson Lab \cite{E94}.
\begin{figure}
\begin{center}
\leavevmode {\epsfysize= 14cm \epsffile{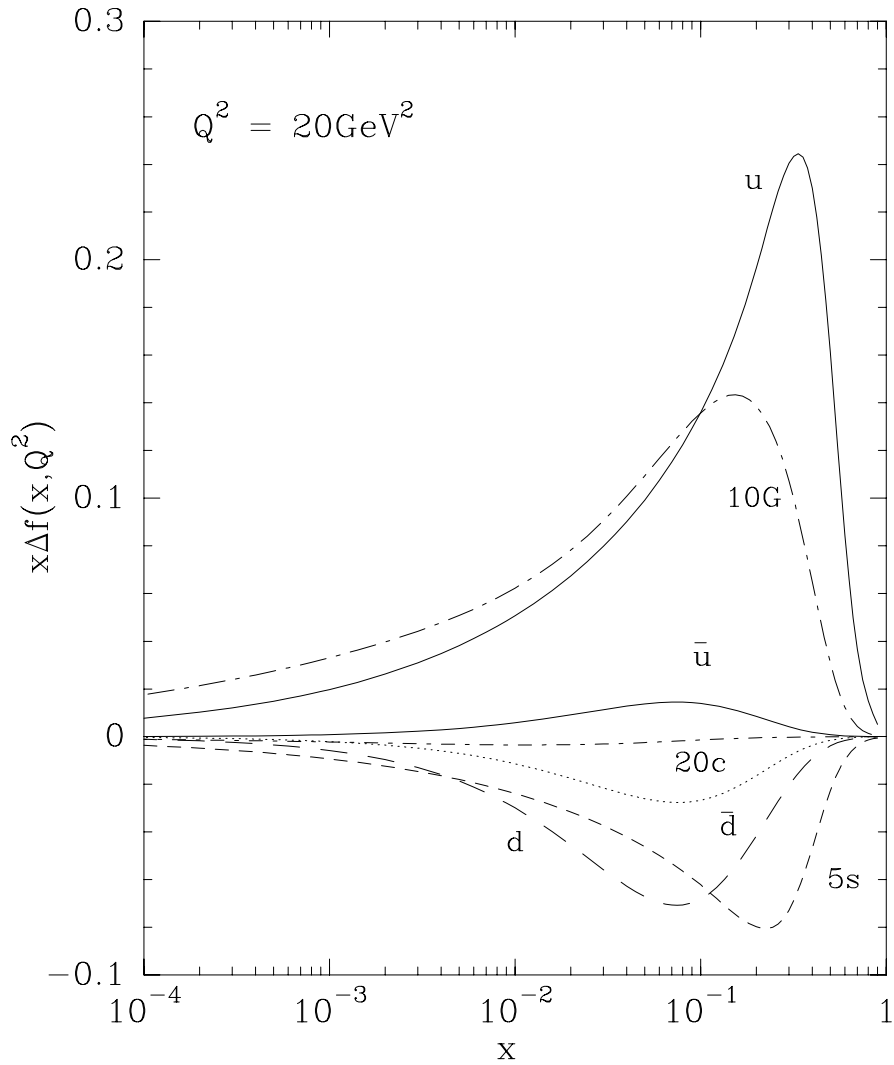}}
\end{center}
\caption[*]{\baselineskip 1pt
The different polarized parton distributions ($f=$$u$,~$d$,~$\bar
u$,~$\bar d$,~$s (\bar s)$,~$c$ and $G$) after NLO evolution,
at $Q^2=20\mbox{GeV}^2$, as a function of $x$.}
\label{Fig18}
\end{figure}
\clearpage
\newpage

\begin{figure}
\begin{center}
\leavevmode {\epsfysize= 14cm \epsffile{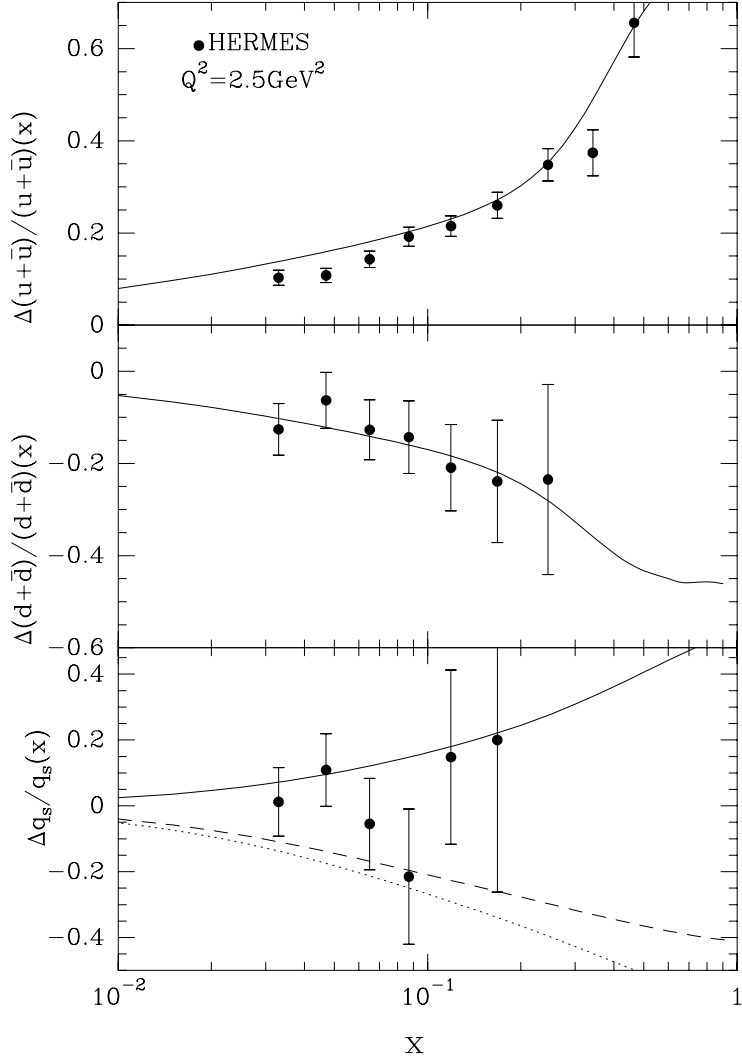}}
\end{center}
\caption[*]{\baselineskip 1pt
HERMES data on $(\Delta u + \Delta \bar u)/ (u +\bar u),
{}~(\Delta d + \Delta \bar d)/( d +\bar d), {}~\Delta q_s / q_s$ as function 
of $x$ at fixed $Q^2 = 2.5\mbox{GeV}^2$ \cite{hermes}. The curves are our model
calculations. For the sea quarks $\Delta \bar u / \bar u$ (solid curve)
,~ $\Delta \bar d / \bar d$ (dashed curve)  and  $\Delta s / s$
(dotted curve).}
\label{Fig19}
\end{figure}
\clearpage
\newpage

\begin{figure}
\begin{center}
\leavevmode {\epsfysize= 14cm \epsffile{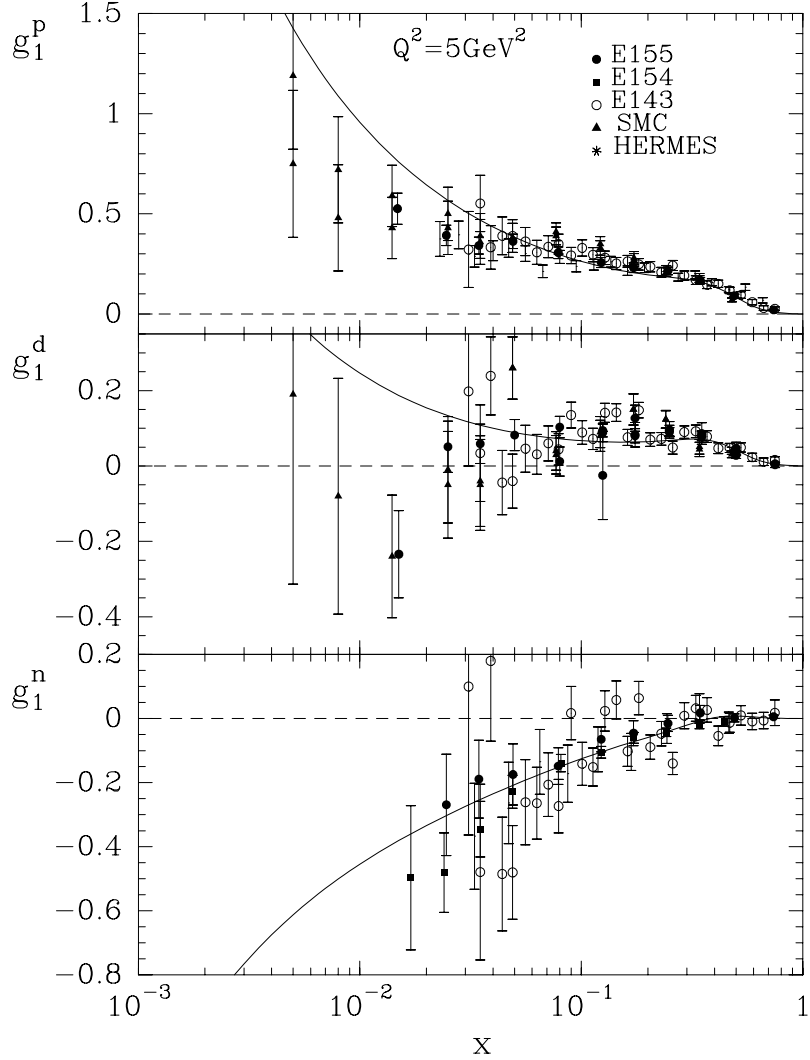}}
\end{center} 
\caption[*]{\baselineskip 1pt
$g_1^{p,d,n}(x,Q^2)$ as function of $x$ for different $Q^2$ values,
from E155, E154, E143, SMC, HERMES experiments.
The curves correspond to our model predictions at $Q^2 = 5\mbox{GeV}^2$.}
\label{Fig20}
\end{figure} 
\clearpage
\newpage
 
\begin{figure}
\begin{center}
\leavevmode {\epsfysize= 14cm \epsffile{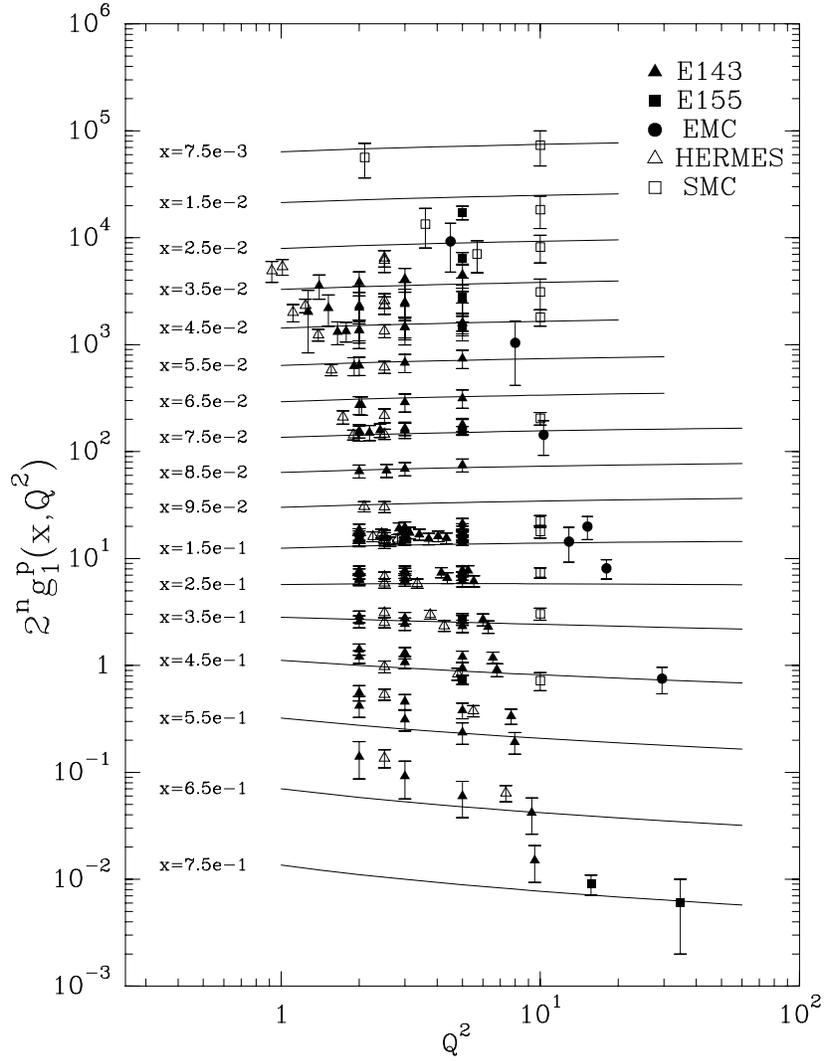}}
\end{center}
\caption[*]{\baselineskip 1pt
$2^ng_1^p(x,Q^2)$ as function of $Q^2$ for different $x$ values.
 $n = 0$ corresponds to $x=0.75$ and $n = 16$ to $x = 7.5~10^{-3}$. 
Experimental data are rebined to the nearest $x$ values.}
\label{Fig21}
\end{figure}
\clearpage
\newpage

\begin{figure}
\begin{center}
\leavevmode {\epsfysize= 14cm \epsffile{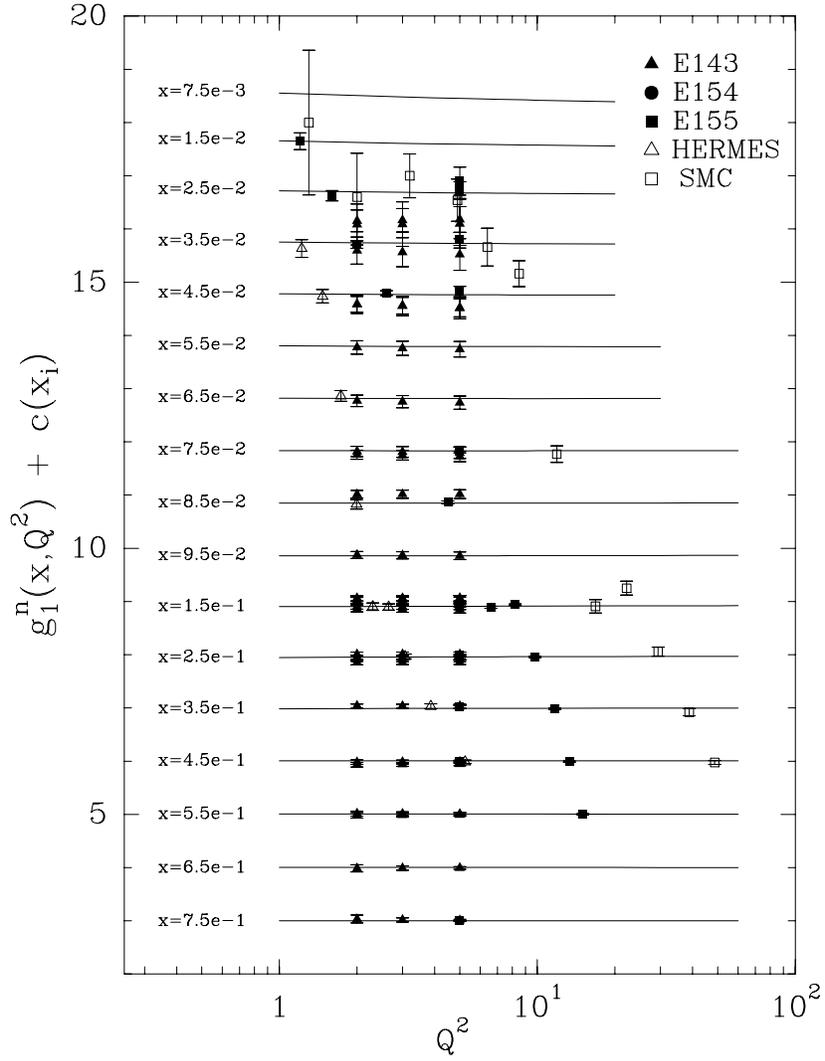}}
\end{center}
\caption[*]{\baselineskip 1pt
$g_1^n(x,Q^2)$ as function of $Q^2$ for different $x$ values.
The function $c(x_i) = 19 -i$, $i = 0$ corresponds to $x = 7.5~10^{-3}$.
Experimental data are rebined to the nearest $x$ values.}
\label{Fig22}
\end{figure}
\clearpage
\newpage

\begin{figure}
\begin{center}
\leavevmode {\epsfysize= 14cm \epsffile{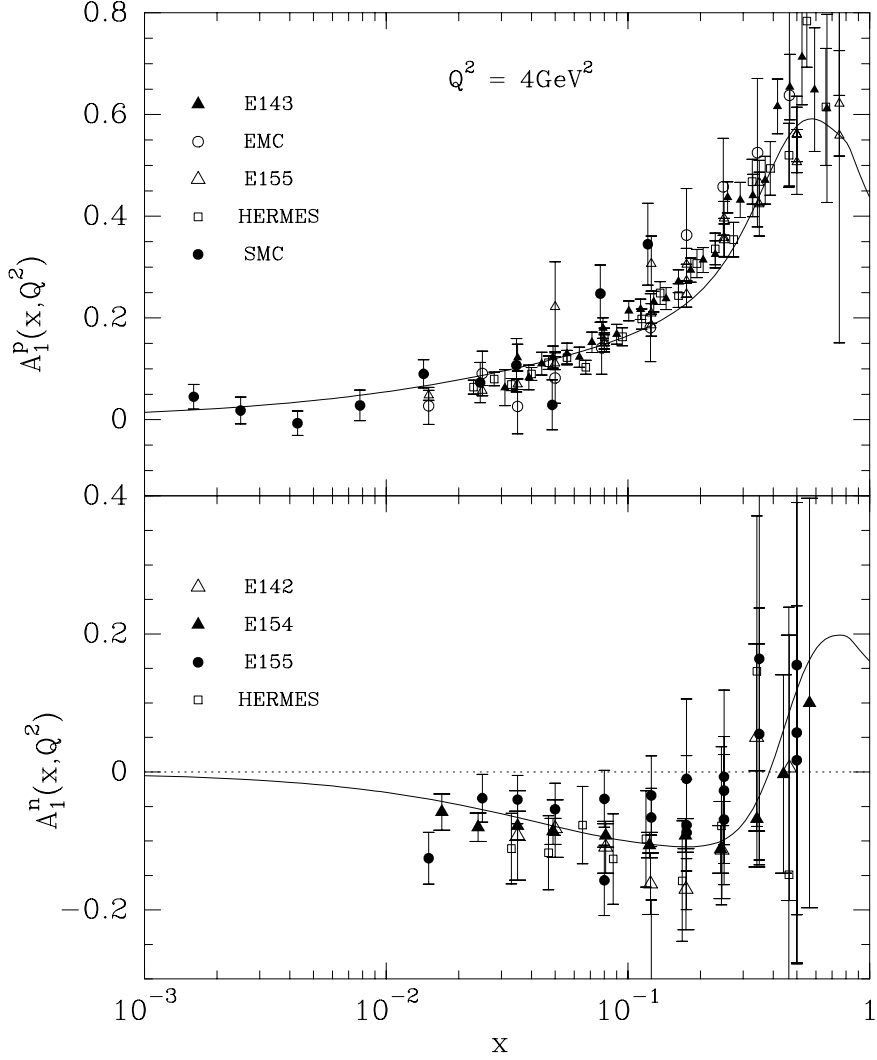}}
\end{center}
\caption[*]{\baselineskip 1pt
Compilation of the asymmetries $A_1^p$ and $A_1^n$
from E155, E154, E142, E143, EMC, SMC and HERMES experiments 
\cite{EMC}-\cite{ABEal}.
The curves correspond to our model predictions at $Q^2 = 4\mbox{GeV}^2$.}
\label{Fig23}
\end{figure}
\clearpage
\newpage

\section{Helicity asymmetries for weak boson production at RHIC-BNL}
Next we propose some tests of our PPD in hadronic collisions in the 
framework of the spin program at RHIC-BNL \cite{BSSW}. As we have seen in
Section 3, the production of $W^{\pm}$ in $pp$ collisions is very relevant
to probe the behavior of the $\bar d(x) / \bar u(x)$ ratio and since $W$
bosons are produced through a pure $V-A$ interaction, they are also
an ideal tool to study the spin structure of the nucleon.

Let us consider the parity-violating helicity asymmetry $A^{PV}_L(W)$
defined as
\begin{equation}
A_{L}^{PV}(W) = 
{ d\sigma^W_-/dy - d\sigma^W_+/dy \over d\sigma^W_-/dy + d\sigma^W_+/dy}~,
\label{28}
\end{equation}
where $\pm$ stands for the helicity of one polarized proton beam. For $W^+$, 
 at the lowest order of the Drell-Yan production mechanism, it reads \cite{BSd}
\begin{equation}
A_{L}^{PV}(W^+) =
{\Delta u(x_a, M^2_W) \bar d(x_b, M^2_W) -\Delta \bar d(x_a, M^2_W)
u(x_b, M^2_W) \over u(x_a, M^2_W) \bar d(x_b, M^2_W)+ \bar d(x_a, M^2_W)
u(x_b, M^2_W)}~,
\label{29}
\end{equation}
assuming the proton $a$ is polarized. Here $x_a,x_b$ are defined as in 
Eq. (\ref{24})
and for $W^-$ production one interchanges the quark flavor $u$ and $d$. 
The calculation of these asymmetries is therefore very simple and the results 
using our PPD are 
presented in Fig. 24  at $\sqrt s = 350\mbox{GeV}$ and $ 500\mbox{GeV}$. 
The asymmetries
decrease for increasing energy and we recall that higher-order corrections
have very small effects on these predictions \cite{KaGeh}. 
The general trend of $A_{L}^{PV}(W)$ 
can be easily understood and, for example at $\sqrt s = 500\mbox{GeV}$  
near $y=+1$, $A_{L}^{PV}(W^+) \sim \Delta u /u$ and
$A_{L}^{PV}(W^-) \sim \Delta d /d$, evaluated at $x=0.435$. 
Similarly for near $y=-1$, $A_{L}^{PV}(W^+) \sim -\Delta \bar d /\bar d$ and
$A_{L}^{PV}(W^-) \sim -\Delta \bar u /\bar u$, evaluated at $x=0.059$. 
Given the expected rates
for $W^{\pm}$ production at RHIC-BNL and the high degree of the 
proton beam polarization \cite{BSSW}, it will
be possible to check these predictions of the statistical approach to 
a high accuracy, in 
particular for the flavor separation of the antiquarks polarized distributions.
It will provide the first reliable determination of $\Delta \bar u$ and 
$\Delta \bar d$.
One can also consider the asymmetry $A_L^{PV}$ for the $Z^0$ production 
whose expression
is given in Ref.~\cite{BSd}. Our prediction for this asymmetry is displayed 
in Fig. 25, but
since the $Z^0$ is not a pure left-handed object the interpretation of 
the result in terms of the PPD is less obvious than in the $W^{\pm}$ case. 
Moreover the $Z^0$ production rate at RHIC-BNL
will be less copious, so the expected precision will be reduced.

In $pp$ collisions for $W^{\pm}$ production where both protons beams 
are polarized, there is another observable
which is sensitive to the antiquark polarized distributions, that is 
the parity-conserving double helicity asymmetry $A_{LL}^{PC}(W)$ defined as
\begin{equation}
A_{LL}^{PC}(W) = 
{ d\sigma^W_{++}/dy + d\sigma^W_{--}/dy - d\sigma^W_{+-}/dy 
- d\sigma^W_{-+}/dy \over
d\sigma^W_{++}/dy + d\sigma^W_{--}/dy + d\sigma^W_{+-}/dy 
+ d\sigma^W_{-+}/dy}~.
\label{30}
\end{equation}
For $W^+$ production it reads, at the lowest order, \cite{BSc}
\begin{equation}
A_{LL}^{PC}(W^+) = -
{\Delta u(x_a,M^2_W) \Delta \bar d(x_b,M^2_W) 
+ \Delta \bar d(x_a,M^2_W)\Delta u(x_b,M^2_W)
\over u(x_a,M^2_W)\bar d(x_b,M^2_W) +\bar d(x_a,M^2_W) u(x_b,M^2_W)}~. 
\label{31}
\end{equation}
We simply notice that if the antiquarks are unpolarized, {\it i.e.} 
$\Delta \bar q_i(x)=0$,
it leads immediately to $A_{LL}^{PC}(W^+)=0$. 
Clearly $A_{LL}^{PC}(W^+)$ is symmetric
around $y=0$ and one obtains $A_{LL}^{PC}(W^-)$ by interchanging the 
quark flavor $u$ and $d$.
Given the signs of the PPD we have obtained in the statistical approach, 
it is obvious that
$A_{LL}^{PC}(W^{\pm})$ are positive in the $y$ range we are considering, 
as shown in Fig. 26.
The asymmetries decrease for increasing energy but they remain sizeable 
even at the highest RHIC-BNL energy.

We have also calculated this asymmetry for $Z^0$ production and 
the results are 
presented in Fig. 27. The sign change with respect to $A_{LL}^{PC}(W^{\pm})$ 
is due to the fact that the expression of $A_{LL}^{PC}(Z^0)$
\cite{BSc} is driven by terms of the type 
$ - \Delta u(x_a) \Delta \bar u(x_b)$ or 
$ - \Delta d(x_a) \Delta \bar d(x_b)$, which are both negative in the 
statistical approach.
 
Finally let us mention briefly a realistic possibility of having polarized 
${}^3He$ at 
RHIC-BNL, which allows to consider $pn$ collisions with polarized neutrons. 
The same asymmetries
can be calculated for this case and, in particular, we have checked that 
$A_L^{PV}(W^{\pm})$ for
$p \overrightarrow n$ collisions is directly obtained from 
$A_L^{PV}(W^{\pm})$ for 
$p \overrightarrow p $ collisions, by exchanging $W^+$ and $W^-$,
as a consequence of isospin symmetry. For $ \overrightarrow p n $ 
collisions with only the proton beam polarized, the results are very 
close to those obtained for $\overrightarrow p p$ collisions.
\begin{figure}
\begin{center}
\leavevmode {\epsfysize= 14cm \epsffile{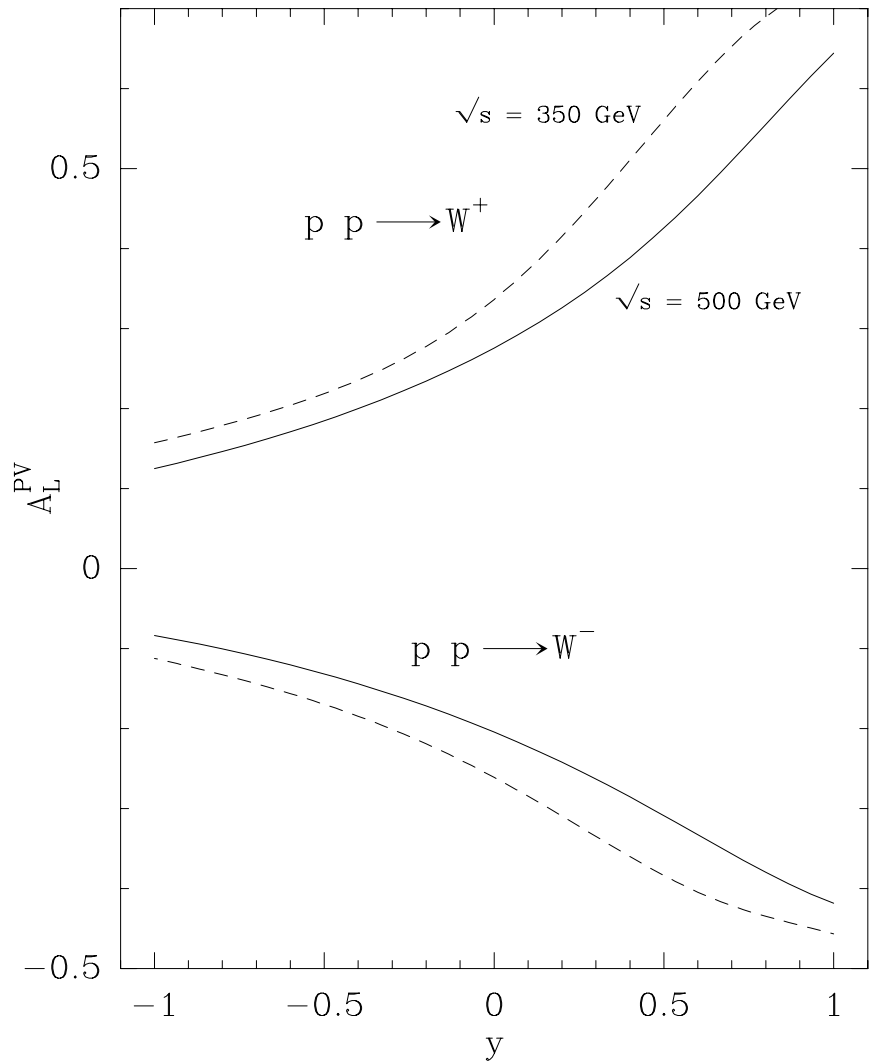}}
\end{center}
\caption[*]{\baselineskip 1pt
The parity violating asymmetry $A_L^{PV}$
for $ p p \rightarrow W^{\pm}$ production versus the $W$ rapidity
at $\sqrt{s} = 350 \mbox{GeV}$ (dashed curve)
and $\sqrt{s} = 500 \mbox{GeV}$ (solid curve).}
\label{Fig24}
\end{figure}

\begin{figure}
\begin{center}
\leavevmode {\epsfysize= 14cm \epsffile{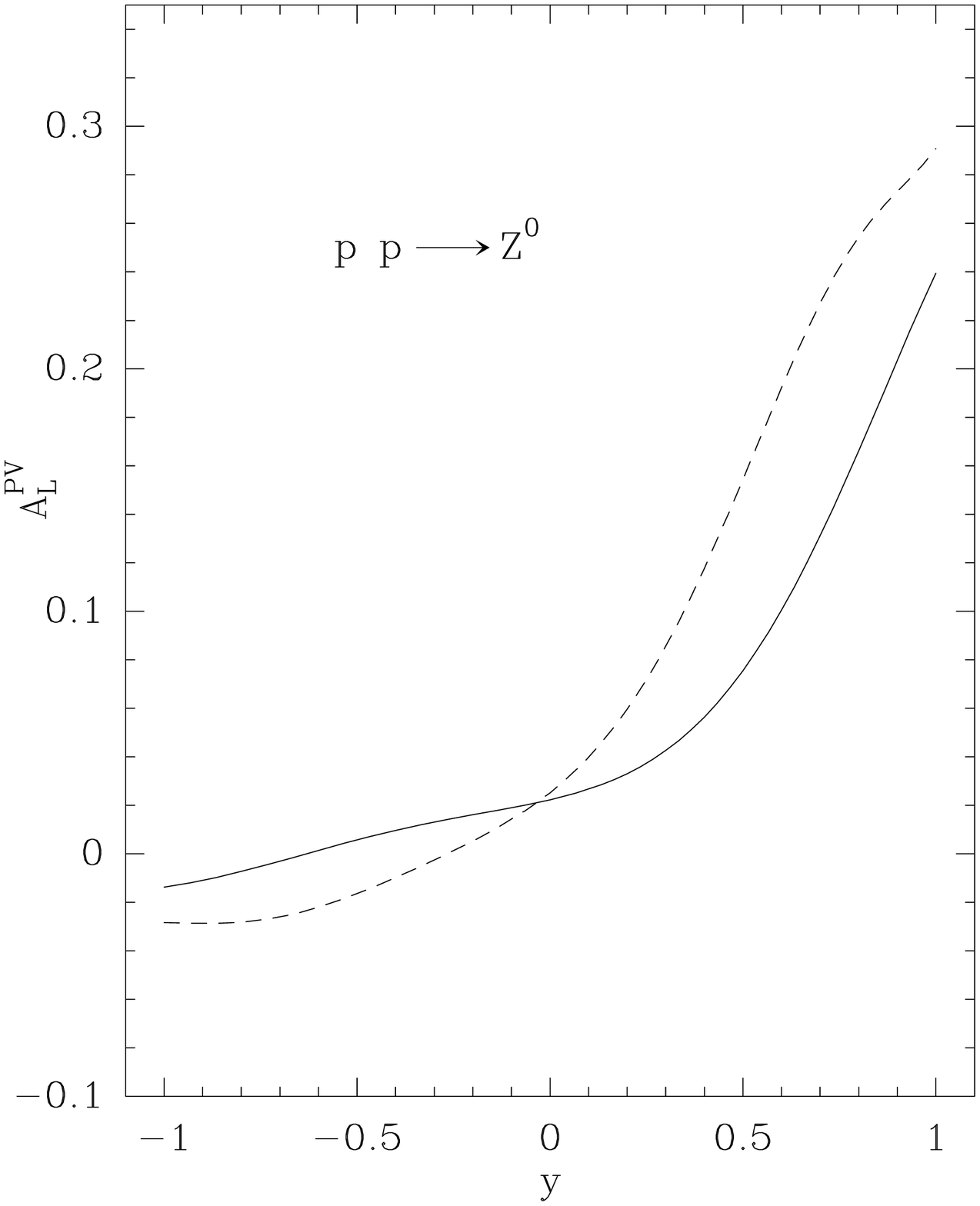}}
\end{center}
\caption[*]{\baselineskip 1pt
The parity violating asymmetry $A_L^{PV}$
for $ p p \rightarrow Z^0$ production versus the $Z^0$ rapidity
at $\sqrt{s} = 350 \mbox{GeV}$ (dashed curve)
and $\sqrt{s} = 500 \mbox{GeV}$ (solid curve).}
\label{Fig25}
\end{figure}

\begin{figure}
\begin{center}
\leavevmode {\epsfysize= 14cm \epsffile{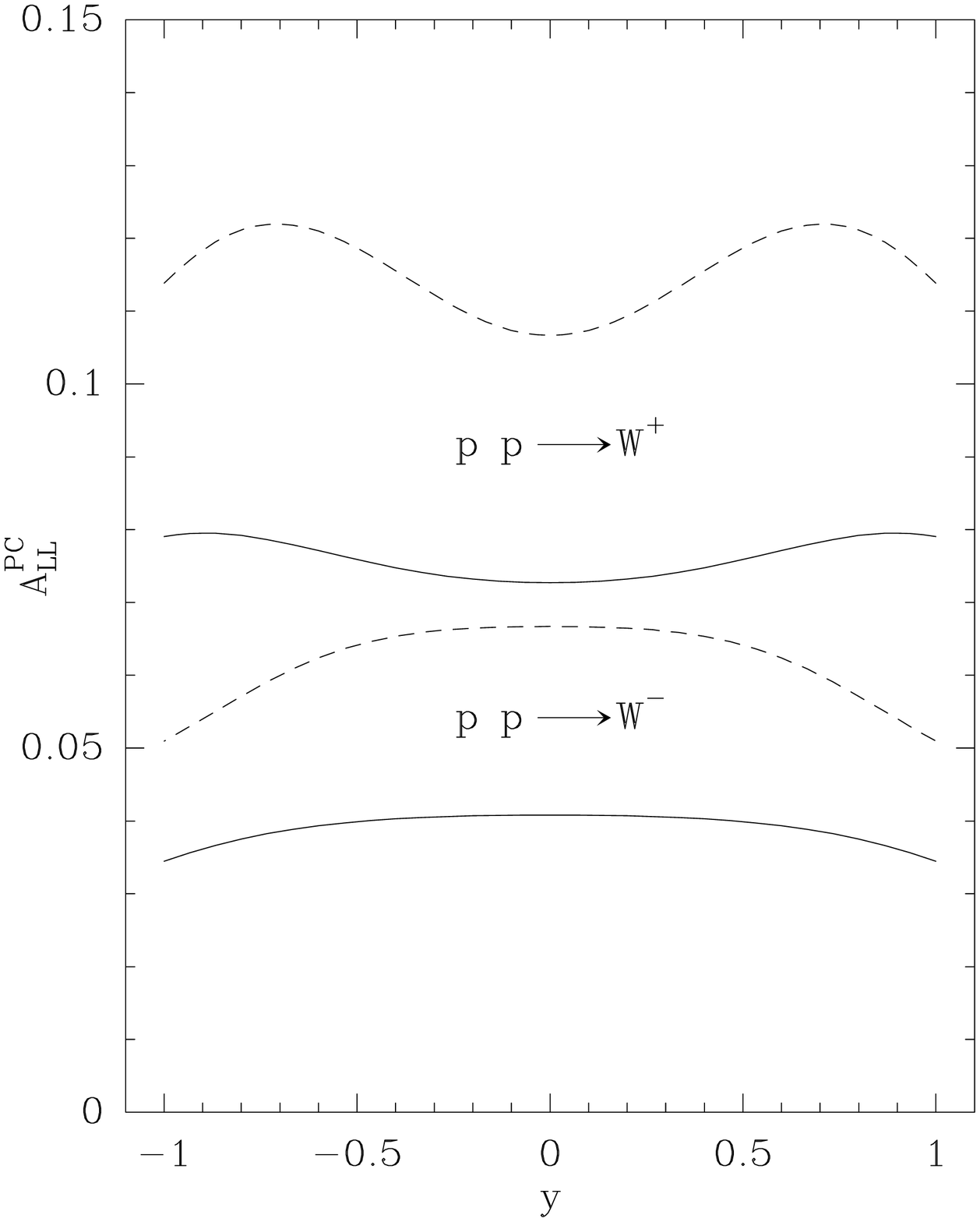}}
\end{center}
\caption[*]{\baselineskip 1pt
Parity conserving double helicity asymmetry $A_{LL}^{PC}$
for $ p p \rightarrow W^{\pm}$ production versus the $W$ rapidity
at $\sqrt{s} = 350 \mbox{GeV}$ (dashed curve)
and $\sqrt{s} = 500 \mbox{GeV}$ (solid curve).} 
\label{Fig26}
\end{figure}

\begin{figure}
\begin{center}
\leavevmode {\epsfysize= 14cm \epsffile{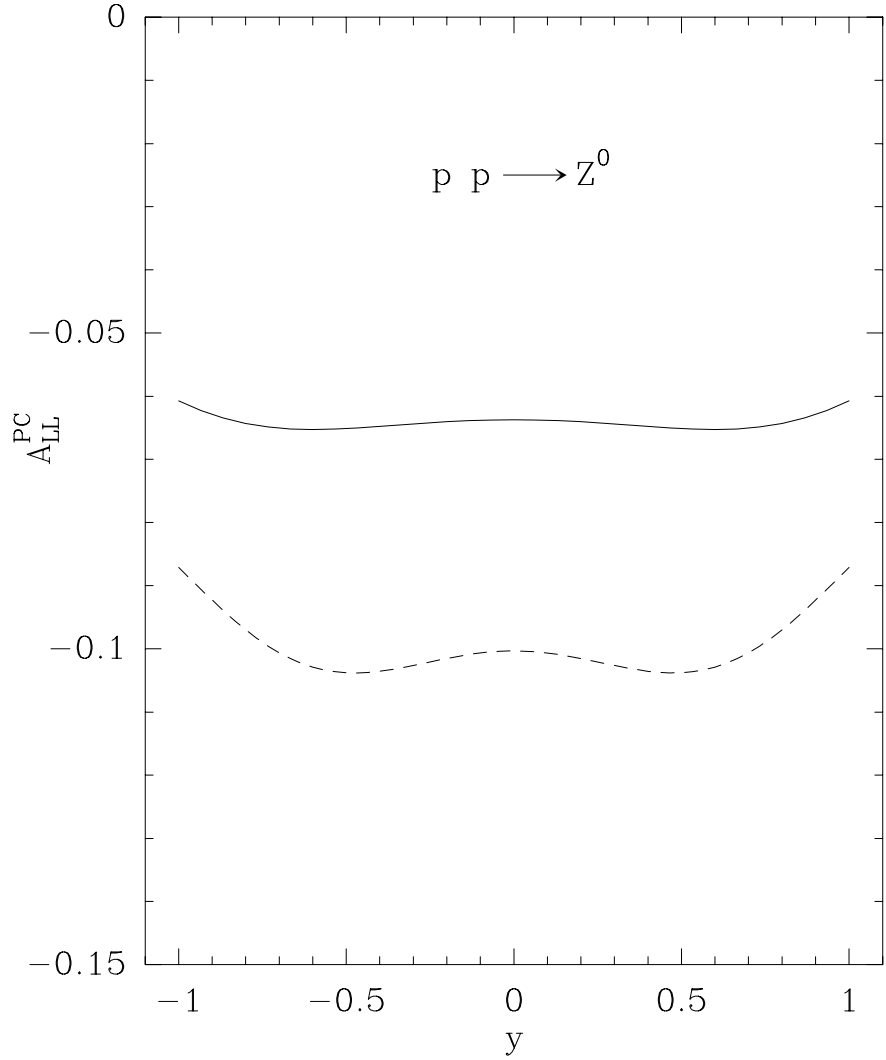}}
\end{center}
\caption[*]{\baselineskip 1pt
Parity conserving double helicity asymmetry $A_{LL}^{PC}$
for $ p p \rightarrow Z^0$ production versus the $Z^0$ rapidity
at $\sqrt{s} = 350 \mbox{GeV}$ (dashed curve)
and $\sqrt{s} = 500 \mbox{GeV}$ (solid curve).}
\label{Fig27}
\end{figure}
\clearpage
\newpage
\section{Concluding remarks}
We have constructed a new set of polarized parton distributions 
in the framework of a statistical physical picture of the nucleon. 
The chiral properties of QCD
lead to simple relations between quark and antiquark distributions. 
We have obtained a good description of all unpolarized
and polarized structure functions measured in DIS,
$F_2^{p,d}(x,Q^2)$, $xF_3^{\nu N}(x,Q^2)$ and $g_1^{p,d,n}(x,Q^2)$, in a wide
kinematic range, in terms of {\it eight} free
parameters, a small number which could be reduced even further. We have also
proposed a simple expression for the gluon distribution $G(x,Q_0^2)$,
at the input scale $Q_0^2$, with
no additional free parameter.
In view of the poor present knowledge of the polarized gluon distribution
we took $\Delta G(x,Q_0^2)=0$. Our analysis 
of the data was done by using a DGLAP $Q^2$ evolution at NLO, to get 
access to a broad kinematic range of $x$ and $Q^2$. 
This approach predicts specific flavor
symmetry breaking of the light antiquarks. First the inequality 
$\bar d(x,Q^2) > \bar u(x,Q^2)$, 
a natural consequence of the Pauli principle, which is 
questioned for $x>0.2$ by the E866 experiment \cite{E866}. To answer this 
challenging issue, we propose to measure 
the ratio of the unpolarized cross sections for the
production of $W^+$ and $W^-$ in $pp$ collisions
at RHIC-BNL, for which we have a definite prediction. Second, this 
approach leads to $\Delta \bar u >0$ and $\Delta \bar d <0$, so these first
moments of the light antiquarks distributions give a positive contribution
to the Bjorken sum rule. The precise tests for the $x$ and $Q^2$ dependence
of these distributions will also be done at RHIC-BNL, by the 
measurements of the
helicity asymmetries in the $W^{\pm}$ and $Z^0$ production, 
which we have calculated.
The strange quarks unpolarized and polarized distributions, 
are constructed by means of
empirical expressions in terms of the light antiquark distributions, but this
should be improved in the future with some more fundamental arguments.
We have also stressed the relevance of the large $x$ region for the light quark
distributions and the behavior of our distributions can be tested in current
experiments at Jefferson Lab. Although we still miss a deep understanding
of all the features of this approach, as well as a clear interpretation 
of some of
the outgoing parameters, we think it is very promising and challenging in view
of the future tests we have identified.
\clearpage
\newpage
\noindent {\bf Acknowledgments}\\
F. Buccella wishes to thank, for warm hospitality, the Centre de Physique
Th\'eorique where part of this work was done.
We thank T. Gehrmann for providing us his code used in Ref. \cite{KaGeh}
for NLO $Q^2$ evolution of the helicity asymmetries. We also 
thank Z.-E. Meziani, 
N.Saito, A.W. Thomas and W.Vogelsang for suggestions and fruitful discussions.
\clearpage
\newpage


\end{document}